\documentclass[10pt]{bmc_article}

\usepackage{cite} % Make references as [1-4], not [1,2,3,4]
\usepackage{url}  % Formatting web addresses
\usepackage{ifthen}  % Conditional
\usepackage{multicol}   %Columns
\usepackage[utf8]{inputenc} %unicode support
\usepackage[numbers,sort&compress]{natbib}

\usepackage{upgreek}
\usepackage{amssymb}
\usepackage{amsmath}
\usepackage{array}

\newcommand{\tabincell}[2]{\begin{tabular}{@{}#1@{}}#2\end{tabular}}

\usepackage{graphicx}
\usepackage[caption=false,font=footnotesize]{subfig}

\newboolean{publ}

\newenvironment{bmcformat}{\baselineskip20pt\sloppy\setboolean{publ}{false}}{\baselineskip20pt\sloppy}

\begin{document}
\begin{bmcformat}

\bibliographystyle{bmc_article}


\title{Adaptive Matching Pursuit for Off-grid Compressed Sensing\footnote{This article was published in http://asp.eurasipjournals.com/content/2012/1/76, titled as Adaptive Matching Pursuit with Constrained Total Least Squares.}}

%%%%%%%%%%%%%%%%%%%%%%%%%%%%%%%%%%%%%%%%%%%%%%
%%                                          %%
%% Enter the authors here                   %%
%%                                          %%
%% Ensure \and is entered between all but   %%
%% the last two authors. This will be       %%
%% replaced by a comma in the final article %%
%%                                          %%
%% Ensure there are no trailing spaces at   %%
%% the ends of the lines                    %%
%%                                          %%
%%%%%%%%%%%%%%%%%%%%%%%%%%%%%%%%%%%%%%%%%%%%%%

\author{
Tianyao Huang,%
\email{Tianyao Huang - huangty09@mails.tsinghua.edu.cn}
Yimin Liu\correspondingauthor,%
         \email{Yimin Liu\correspondingauthor - yiminliu@tsinghua.edu.cn}
Huadong Meng%
 \email{Huadong Meng - menghd@tsinghua.edu.cn}
and Xiqin Wang%
\email{Xiqin Wang - wangxq\_ee@tsinghua.edu.cn}%
      }

\address{%
 Department of Electronic Engineering, Tsinghua University,
Beijing, China
}%

\maketitle

%%%%%%%%%%%%%%%%%%%%%%%%%%%%%%%%%%%%%%%%%%%%%%
%%                                          %%
%% The Abstract begins here                 %%
%%                                          %%
%% Please refer to the Instructions for     %%
%% authors on http://www.biomedcentral.com  %%
%% and include the section headings         %%
%% accordingly for your article type.       %%
%%                                          %%
%%%%%%%%%%%%%%%%%%%%%%%%%%%%%%%%%%%%%%%%%%%%%%

\begin{abstract}
Compressive Sensing (CS) can effectively recover a signal
when it is sparse in some discrete atoms. However, in some
applications, signals are sparse in a continuous parameter
space, e.g. frequency space, rather than discrete atoms. Usually,
we divide the continuous parameter into finite discrete grid
points and build a dictionary from these grid points.
However, the actual targets may not exactly lie on the grid
points no matter how densely the parameter is grided, which introduces mismatch between the predefined dictionary and the actual one.
In this paper, a novel method, namely Adaptive Matching Pursuit
with Constrained Total Least Squares (AMP-CTLS), is
proposed to find actual atoms even if they are not included
in the initial dictionary. In
AMP-CTLS, the grid and the dictionary are adaptively updated
to better agree with measurements. The convergence of the algorithm
is discussed, and numerical experiments demonstrate the
advantages of AMP-CTLS.
\end{abstract}

\ifthenelse{\boolean{publ}}{\begin{multicols}{2}}{}

%%%%%%%%%%%%%%%%%%%%%%%%%%%%%%%%%%%%%%%%%%%%%%
%%                                          %%
%% The Main Body begins here                %%
%%                                          %%
%% Please refer to the instructions for     %%
%% authors on:                              %%
%% http://www.biomedcentral.com/info/authors%%
%% and include the section headings         %%
%% accordingly for your article type.       %%
%%                                          %%
%% See the Results and Discussion section   %%
%% for details on how to create sub-sections%%
%%                                          %%
%% use \cite{...} to cite references        %%
%%  \cite{koon} and                         %%
%%  \cite{oreg,khar,zvai,xjon,schn,pond}    %%
%%  \nocite{smith,marg,hunn,advi,koha,mouse}%%
%%                                          %%
%%%%%%%%%%%%%%%%%%%%%%%%%%%%%%%%%%%%%%%%%%%%%%

%%%%%%%%%%%%%%%%
%% Background %%
%%
\section{Introduction}

A new class of techniques called Compressed Sampling or Compressive
Sensing (CS) has been widely used recently, due to the fact that CS techniques have shown good performance in different areas such as signal processing,
communication and statistics; see, e.g., \cite{ref:CS}. Generally, CS
finds the sparsest vector \textbf{x} from measurements
$\textbf{y}=\boldsymbol{\Phi}\textbf{x}$, where $\boldsymbol{\Phi}$
is often referred to as \textit{dictionary} with more columns than
rows, and each column of the dictionary is called an \textit{atom} or a \textit{basis}.\par

Matching Pursuit (MP) is a set of popular greedy approaches to
Compressive Sensing. The basic idea is to sequentially find the
support set of \textbf{x} and then project on the selected atoms.
The atoms selected in the support set are mainly determined by
correlations between atoms and the regularized measurements
\cite{ref:SP}. MP methods include standard MP \cite{ref:MP},
and several other examples, such as Orthogonal Matching Pursuit (OMP) \cite{ref:OMP1}, Regularized OMP (ROMP) \cite{ref:ROMP}, Stage-wise OMP (StOMP) \cite{ref:StOMP}, Compressive Sampling Matching Pursuit (CoSaMP) \cite{ref:CoSaMP} and Subspace Pursuit (SP) \cite{ref:SP}. \par

These MP methods \cite{ref:SP,ref:MP,ref:OMP1,ref:CoSaMP,ref:ROMP,ref:StOMP} do not consider the off-grid problem in grid-based CS approaches. In some applications of CS, such as harmonic retrieval and radar signal processing (e.g. range profiling \cite{ref:RSFR,ref:SFR}, direction of arrival estimate
\cite{ref:DOA1,Zheng2010,Zheng2011}), we usually divide a continuous parameter space into discrete grid points to generate the dictionary. For example, in harmonic retrieval, frequency space is divided and dictionary is a discrete Fourier transform (DFT) matrix. The off-grid problem emerges when the actual frequencies are placed off the predefined grid. The mismatch between the predefined and actual atom can lead to performance degradation in sparse recovery (e.g., \cite{ref:Yuejie2,ref:ZhuHao,ref:mis1}).\par

The grid misalignment problem in CS has recently received growing interest.
The sensitivity of CS to the mismatch between the predefined and actual atoms is studied in \cite{ref:Yuejie2}; however, the focus of that paper is mainly on mismatch analysis rather than development of an algorithm. Cabrera et al. \cite{ref:insert} and Zhu et al. \cite{ref:ZhuHao} respectively provided an Iterative Re-Weighted (IRW)-based and a Lasso-based method to recover an unknown vector considering the atom misalignment, whereas we focus on MP methods in this paper. Compared with IRW or Lasso, MP methods greedily find the support set and greatly reduce the dimension of the CS problem; thus, they have an advantage in computability. Gabriel \cite{ref:tree} proposed best basis compressive sensing in a tree-structured dictionary, but some dictionaries (e.g., DFT matrix) do not possess a tree structure.\par

To alleviate the off-grid problem in Matching Pursuit, we developed Adaptive Matching Pursuit with Constrained Total Least Squares (AMP-CTLS). In AMP-CTLS, we model the grid as an unknown parameter, and adaptively search for the best one. We choose harmonic retrieval to demonstrate the performance of AMP-CTLS. The algorithm can also be applied to jointly estimate range and velocity in randomized step frequency (RSF) radar. Note that in the RSF scenario range-velocity estimation is hard to be directly solved by subspace-based methods, e.g. Capon's method, MUSIC and ESPRIT \cite{ref:stoica2005spectral}. Since only one snapshot data is available in RSF radar, to obtain the covariance matrix these subspace-based methods need to apply smoothing method, which requires uniform and linear condition \cite{ref:stoica2005spectral}. However, this condition is not satisfied in the case of random frequency model.

This paper is structured as follows. Section \ref{sec:SigMod} introduces
grid-based CS and outlines the procedures of AMP-CTLS. In Section \ref{sec:HarRet} and \ref{sec:RSF}, we discuss the implementation of AMP-CTLS in harmonic retrieval and RSF radar, respectively. In Section \ref{sec:sim}, numerical examples are presented to illustrate merits of AMP-CTLS. Section \ref{sec:conlusion} is dedicated to a brief conclusion. \par

Notations: $(\cdot)^\text{H}$ denotes conjugate transpose matrix;
$(\cdot)^\text{T}$ transpose matrix; $(\cdot)^*$ conjugate matrix;
$(\cdot)^\dag$ pseudo-inverse matrix; $\textbf{I}_L/\textbf{0}_L$ the $L\times L$ identity/zero matrix; $\|\cdot\|_2$ the $\ell_2$ norm; $\{\cdot\}$ denotes a set; $\mid\cdot\mid$ the absolute value of a complex number or the
cardinality of a set; $(\cdot)_\Lambda$ denotes elements/columns indexed in the set $\Lambda$ of a vector/matrix; $\text{supp}(\cdot)$ is the support set of a vector, that is, the indices of the nonzero elements in the vector;
$\text{Re}(\cdot)$ the real part of a complex number; $\otimes$
denotes the right Kronecker product \cite{ref:matrix}; and
$E[\cdot]$ denotes the expectation of a random variable.

\section{Grid-based CS and the AMP-CTLS Algorithm}{\label{sec:SigMod}}

The signal model of grid-based CS is introduced in Subsection \ref{subsec:gridCS}. We combine the greedy idea of MP methods and the Constrained Total Least Squares (CTLS) technique \cite{ref:CTLS}, and thus produce AMP-CTLS to alleviate the off-grid problem. In AMP-CTLS, the grid is cast as an unknown parameter, and is jointly estimated together with \textbf{x}. In Subsection \ref{subsec:mainidea}, the framework of AMP-CTLS is given. Subsection \ref{subsec:IJE} is dedicated to the Iterative Joint Estimator (IJE) algorithm, which is implemented in AMP-CTLS. In the IJE algorithm, the CTLS technique is used, which is presented in Subsection \ref{subsec:CTLS}. Subsection \ref{subsec:sketch} summarizes the entire procedure of AMP-CTLS. In Subsection \ref{subsec:convergence}, the convergence of IJE is analyzed.

\subsection{Grid-based CS}{\label{subsec:gridCS}}

CS promises efficient recovery of sparse signals. In many applications, signals are sparse in a continuous parameter space rather than finite discrete atoms. Usually, we divide the continuous parameter into discrete grid points and cast the problem as a grid-based CS model:
\begin{equation}{\label{Equ:matrixmodel}}
    \textbf{y} = {\boldsymbol{\Phi}\left(\textbf{g}\right)} \textbf{x} + \textbf{w},
\end{equation}
where ${\bold y} \in {\mathbb{C}}^{M \times 1}$ and ${\bold w} \in {\mathbb{C}}^{M \times 1}$ are measurement vector and white Gaussian noise (WGN) vector, respectively. ${\bf x} \in {\mathbb{C}}^{N \times 1}$ is to be learned. ${\bold g} \in {\mathbb{C}}^{N \times 1}$ are discrete grid points $\textbf{g}=[g_1,g_2,\dots,g_N]$. $\boldsymbol{\Phi}(\textbf{g}) \in {\mathbb{C}}^{M \times N}$ is built from $\bold g$, $\boldsymbol{\Phi}(\textbf{g}) = \left[\phi(g_1),\phi(g_2),\dots,\phi(g_N) \right]$, and the mapping ${\bf g} \bf \rightarrow {\bf{\Phi}}$ is known.
For example, to recover a frequency sparse signal, we grid the frequency space into discrete frequency points ${\bold g} = [0, \frac{1}{N}, \frac{2}{N}, \dots,\frac{N-1}{N}]^{\rm T}$. $\boldsymbol{\Phi}$ is a DFT matrix, of which the $m$th-row, $n$th-column element is $\exp \left( j2\pi \frac{n}{N} m \right)$. However, the signal is only sparse in the DFT atoms if all of the sinusoids are exactly at the pre-defined grid points \cite{ref:Yuejie2}. In some cases, no matter how densely we grid the frequency space, the sinusoids could be off-grid, which saps the performance of CS methods \cite{ref:Yuejie2}.\par

\subsection{Main Idea of AMP-CTLS}{\label{subsec:mainidea}}

The off-grid problem usually emerges because we do not often have enough priori knowledge to generate a perfect grid to guarantee that all of the signals exactly lie on grid points. Thus, we cast the grid as an unknown parameter, and search for the best grid \textbf{g} as well as the sparsest \textbf{x} by solving the optimum problem:
\begin{equation}{\label{Equ:OptmWhole}}
    \hat{\textbf{x}} , \hat{\textbf{g}}  = \mathop {\arg \min } \limits_{\textbf{x},\textbf{g}} {\left\|\textbf{x}
    \right\|_0},
     s.t. \left\| {\textbf{y} - \boldsymbol{\Phi}(\textbf{g})\textbf{x}} \right\|_2^2 \le \eta,
\end{equation}
where $\eta$ is the noise power. Equ. (\ref{Equ:OptmWhole})
is similar to that used in traditional MP methods \cite{ref:SP,ref:MP,ref:OMP1,ref:CoSaMP,ref:ROMP,ref:StOMP}, except that we
recover \textbf{x} {\it{and}} simultaneously estimate the grid.
In most cases, solving (\ref{Equ:OptmWhole}) is a complex non-linear optimum problem. In this paper, an iterative method is introduced. \par

AMP-CTLS inherits the greedy idea from MP methods, which use correlations to iteratively find the support set. In each iteration, one or more atoms are added into the support set.
Suppose the support set is obtained as $\Lambda ^{(k)}$ after the $k$th iteration, and denote the corresponding grid points as ${\widehat {\textbf{g}}}_{\Lambda}^{\left( k \right)}$. In traditional MP methods \cite{ref:SP,ref:MP,ref:OMP1,ref:CoSaMP,ref:ROMP,ref:StOMP}, $\textbf{x}_{\Lambda}$ is estimated by solving a least squares problem.
In AMP-CTLS, considering the off-grid problem, we jointly search for $\textbf{x}_{\Lambda}$ and the best grid points in the neighboring continuous region of ${\widehat {\textbf{g}}}_{\Lambda}^{\left( k \right)}$ via (\ref{Equ:gridxlambda}), in which we minimize norm of the \textit{residual
error}, which is defined as ${\bf r}={\bf y}-{\bf \Phi}\left( {\textbf{g}_\Lambda} \right){{{\bf x}_\Lambda} }$.
\begin{equation}{\label{Equ:gridxlambda}}
        {{\widehat
        {\bf{x}}}_{\Lambda }^{\left( {k + 1} \right)}} , {{\widehat {\textbf{g}}_{\Lambda}}^{\left( {k + 1} \right)}} = \mathop {\arg \min
        }\limits_{{{\bf{x}_\Lambda}} , {\textbf{g}_\Lambda} } \left\| {{\bf{y}} - {{\bf{\Phi
        }} }\left( {\textbf{g}_\Lambda} \right){{\bf{x}_\Lambda} }} \right\|_2^2
\end{equation} \par

We develop the Iterative Joint Estimator (IJE) algorithm to solve (\ref{Equ:gridxlambda}), which is detailed in ensuing subsection.

\subsection{IJE Algorithm}{\label{subsec:IJE}}
It is difficult to find an analytical solution to (\ref{Equ:gridxlambda}). The IJE algorithm is devised to seek a numerical solution. Given initial grid points ${\widehat {\textbf{g}}}_{\Lambda}{\left( 0 \right)}$,
IJE searches for the best grid points $\textbf{g}_{\Lambda}$
in the neighborhood of ${\widehat {\textbf{g}}}_{\Lambda}{\left( 0 \right)}$.
The mismatch of the grid is denoted as $\Delta \textbf{g}_{\Lambda} = \textbf{g}_{\Lambda} -
{\widehat {\textbf{g}}}_{\Lambda}{\left( 0 \right)} = \left[ {\Delta g _1,\dots,\Delta g_{|\Lambda|}} \right]^{\rm{T}}$.
IJE includes three steps: calculate the estimation of the mismatch, $\widehat {\Delta \textbf{g}}_{\Lambda}$; update the grid with $\widehat {\Delta
\textbf{g}}_{\Lambda}$; and estimate $\textbf{x}_\Lambda$ with projection onto the new grid points.
These three steps are executed iteratively to pursue more accurate results.
To distinguish from iterations in search for the support set in (\ref{Equ:gridxlambda}), we denote \textit{l} as the counter of loops in IJE; thus, IJE is expressed as follows:
\begin{equation}{\label{Equ:cCTLS}}
    {{\widehat {\Delta {\bf{g}}}_{\Lambda }}{\left( {l}
    \right)}},{{\bf{x}}_{{\rm{CTLS}}}} = \mathop {\arg \min
    }\limits_{\Delta {{\bf{g}}_\Lambda },{{\bf{x}}_\Lambda }}
    {C_{{\rm{CTLS}}}},
\end{equation}
\begin{equation}{\label{Equ:updategrid}}
    {{\widehat {\bf{g}}}_{\Lambda }{\left( {l + 1} \right)}} =
    {{\widehat {\bf{g}}}_{\Lambda }{\left( {l} \right)}} +
    {{\widehat {\Delta {\bf{g}}}}_{\Lambda }{\left( {l} \right)}},
\end{equation}
\begin{equation}{\label{Equ:LS}}
    {\widehat {{{\bf{x}}}}_{\Lambda }{\left( {l + 1} \right)}} =
    \mathop {\arg \min }\limits_{{{\bf{x}}_\Lambda }} \left\| {{\bf{y}}
    - {\bf{\Phi }}\Big( {{{\widehat {\bf{g}}}_{\Lambda} {\left( {l
    + 1} \right)}}} \Big){{\bf{x}}_\Lambda }} \right\|_2^2.
\end{equation}
In (\ref{Equ:cCTLS}), CTLS technique is applied to simultaneously search for the mismatch $\Delta {{\bf{g}}_\Lambda }$ and $\bf{x}_\Lambda$, and ${{\widehat {\Delta {\bf{g}}}}_{\Lambda }{\left( {l} \right)}}$ and ${{\bf{x}}_{{\rm{CTLS}}}}$ are the results. ${C_{{\rm{CTLS}}}}$
denotes the penalty function of CTLS, which is detailed in Subsection \ref{subsec:CTLS}. Since (\ref{Equ:LS}) is a linear least squares problem, the closed-form solution is
\begin{equation}{\label{Equ:LSx}}
    {\widehat {{{\bf{x}}}}_{\Lambda }{\left( {l + 1} \right)}}=\bigg({{\bf{\Phi }} }\Big( {{{\widehat {\bf{g}}}_{\Lambda} {\left( {l + 1} \right)}}} \Big)\bigg)^\dag {\bf{y}} = \left(\bf{\Phi}^{\rm{H}} {\bf{\Phi}}\right)^{-1}{\bf{\Phi}}^{\rm{H}}{\bf{y}}.
\end{equation}
The loops are terminated when the norm of residual error is scarcely reduced.

\subsection{CTLS Technique}{\label{subsec:CTLS}}
Traditional MP methods \cite{ref:SP,ref:MP,ref:OMP1,ref:CoSaMP,ref:ROMP,ref:StOMP} apply least squares to calculate amplitudes of ${\bf x}_{\Lambda}$ after finding the support set. When there are off-grid signals, mismatches occur in the dictionary; thus, we replace the least squares model with total least squares (TLS) criterion, which is appropriate to deal with the fitting problem when perturbations exist in both the measurement vector and in the dictionary \cite{golub1980analysis}. Since the dictionary mismatches are constrained by errors of grid points, we introduce the Constrained Total Least Squares (CTLS) technique \cite{ref:CTLS} in AMP-CTLS to jointly estimate the grid point errors and ${\bf x}_{\Lambda}$, i.e. solving (\ref{Equ:cCTLS}).  It has been proved that CTLS is a constrained space state maximum likelihood estimator \cite{ref:CTLS}.\par
Suppose that we obtain the
estimate of grid points as ${\widehat {\textbf{g}}}_{\Lambda}
{\left( {l} \right)}$
after $l$th IJE iteration. Assume that the mismatch
$\Delta\textbf{g}_{\Lambda}$ is significantly small; thus we can
approximate the perfect dictionary $\bf{\Phi}
(\textbf{g}_{\Lambda})$ as a linear combination of
the mismatch $\Delta {\bf g}$ with Taylor expansion:
\begin{equation}{\label{Equ:Taylor}}
    {{\boldsymbol{\Phi }}}\left( {\textbf{g}}_{\Lambda} \right) =
    {{\boldsymbol{\Phi }}}\big( {{{\widehat
    {\textbf{g}}}_{\Lambda}{\left( {l} \right)}}} \big) +
    \sum\limits_{i = 1}^{|\Lambda|} {{{\bf{R}}_i}\big( {{{\widehat
    {\textbf{g}}}_{\Lambda}{\left( {l} \right)}}} \big)\Delta {g_i}}
    + \sum\limits_{i = 1}^{|\Lambda|} {o\left( {\Delta g_i^2} \right)},
\end{equation}
where ${\bf R}_i \in {\mathbb{C}}^{M \times |\Lambda|}$ is
\begin{equation}{\label{Equ:Rmatrix}}
    {{\bf{R}}_i}\big( {{{\widehat {\bf{g}}}_{\Lambda} {\left( {l}
    \right)}}} \big) = {\left. {\frac{{\partial {\bf{\Phi }}\left(
    {{{\bf{g}}_\Lambda }} \right)}}{{\partial {g_i}}}}
    \right|_{{{\bf{g}}_\Lambda } = {{{\widehat {\bf{g}}}_{\Lambda
    }{\left( {l} \right)}}}}}
\end{equation}
and $o (\cdot )$ denotes higher order terms.  For simplicity, in this subsection we ignore the iteration counter in the
notations, and ${{\bf{R}}_i}\big( {{{\widehat {\bf{g}}}_{\Lambda
}{\left( {l} \right)}}} \big)$, ${\bf{\Phi }}\big(
{{{\widehat {\bf{g}}}_{\Lambda} {\left( {l} \right)}}} \big)$
are respectively simplified as ${\bf{R}}_i$, $\bf{\Phi}_\Lambda$. Neglect $o\left( {\Delta
g_i^2} \right)$ and the signal model in (\ref{Equ:matrixmodel}) is replaced by:
\begin{equation}{\label{Equ:modelwithmismatch}}
    \textbf{y} = \left(\boldsymbol{\Phi _{\Lambda}}+
    \sum\limits_{i = 1}^{|\Lambda|} {{{\bf{R}}_i}\Delta {g_i}}
    \right) \textbf{x}_{\Lambda} + \textbf{w}.
\end{equation}

CTLS models $\Delta \bf{g} _\Lambda$ as an unknown random perturbation vector. The grid misalignment and the noise vector are combined into a $(M+|\Lambda|)$-dimensional vector ${\bf{v}} = {\left[ {{{\left( {\Delta {{\bf{g}}_\Lambda }} \right)}^{\rm{T}}},{{\bf{w}}^{\rm{T}}}} \right]^{\rm{T}}}$, and CTLS aims at minimizing $\left\| {\bf{v}} \right\|_2^2$. It has been proved that CTLS is a constrained space state maximum likelihood estimator if \textbf{v} is a WGN vector \cite{ref:CTLS}. Thus, we first whiten \textbf{v}. Assume that $\Delta \bf{g}
_\Lambda$ is independent of \textbf{w}. The covariance matrix of
$\Delta \bf{g} _\Lambda$ is ${{\bf{C_g}}} = E\left[
{\Delta {{\bf{g}}_\Lambda }{{\left( {\Delta {{\bf{g}}_\Lambda }}
\right)}^{\rm{H}}}} \right] \in {\mathbb{C}}^{|\Lambda| \times |\Lambda|}$. ${\bf D} \in {\mathbb{C}}^{|\Lambda| \times |\Lambda|}$ obeys ${\bf{C}}_{\bf g}^{ - 1} = {{\bf{D}}^{\rm{H}}}{\bf{D}}$. The variance of white noise \textbf{w} is $\sigma_{\bf{w}}^2$. We denote an unknown normalized vector ${\bf u} \in {\mathbb{C}}^{(M+|\Lambda|) \times 1}$ as (\ref{Equ:u}); thus,
\textbf{u} is a WGN vector.
\begin{equation}{\label{Equ:u}}
    {\bf{u}} = \left[ {\begin{array}{*{20}{c}}
    {{\bf{D}}{{\Delta \bf{g}}_\Lambda }}\\
    {\frac{1}{{{\sigma _{\bf{w}}}}}{\bf{w}}}
\end{array}} \right]
\end{equation}
\par
Minimize the penalty function ${C_{{\rm{CTLS}}}} = \left\| {\bf{u}}
\right\|_2^2$  and (\ref{Equ:cCTLS}) is detailed as follows:
\begin{equation}{\label{Equ:uCTLS}}
    \widehat {\bf{u}},{{\bf{x}}_{{\rm{CTLS}}}} = \mathop {\arg \min }\limits_{{\bf{u}},{{\bf{x}}_\Lambda }} \left\| {\bf{u}} \right\|_2^2,
\end{equation}
\begin{equation}{\label{Equ:uCTLS2}}
    s.t. - {\bf{y}} + \left( {{{\bf{\Phi }}_\Lambda } + \sum\limits_i^{\mid \Lambda \mid} {{{\bf{R}}_i}\Delta {g_i}} } \right){{\bf{x}}_\Lambda } + {\bf{w}} = {\bf{0}}.
\end{equation}
The constraint condition (\ref{Equ:uCTLS2}) can be rewritten as:
\begin{equation}{\label{Equ:Wx}}
    s.t. - {\bf{y}} + {{\bf{\Phi }}_\Lambda }{{\bf{x}}_\Lambda } + {{\bf{W}}_{\bf{x}}}{\bf{u}} = {\bf{0}},
\end{equation}
where ${{\bf{W}}_{\bf{x}}}=\left[ {\bf{H}} \ \ {{\sigma _{\bf{w}}}{{\bf{I}}_M}} \right] \in {\mathbb{C}}^{M \times (|\Lambda|+M)}$. ${\bf H} \in {\mathbb{C}}^{M \times |\Lambda|}$ is defined as
\begin{equation}{\label{Equ:H}}
{\bf{H}} = {\bf{G}}\left( {{{\bf{D}}^{ - 1}} \otimes {{\bf{I}}_{\mid \Lambda \mid}}} \right)\left( {{{\bf{I}}_{\mid \Lambda \mid}} \otimes {{\bf{x}}_\Lambda }} \right),
\end{equation}
where ${\bf{G}} = \left[ {{{\bf{R}}_1},\dots,{{\bf{R}}_{\mid \Lambda \mid}}} \right] \in {\mathbb{C}}^{M \times |\Lambda|^2}$.
The equivalence between (\ref{Equ:uCTLS2}) and (\ref{Equ:Wx}) is
proved as follows:
\begin{eqnarray}{\label{Equ:Equivalency}}
    \left( {\sum\limits_{i = 1}^{\mid \Lambda \mid} {{{\bf{R}}_i}\Delta {g_i}} } \right){{\bf{x}}_\Lambda }
      &=&{\bf{G}}\left( {\Delta {{\bf{g}}_\Lambda } \otimes {{\bf{I}}_{\mid \Lambda \mid}}} \right){{\bf{x}}_\Lambda } \notag \\
     &=& {\bf{G}}\left( {{{\bf{D}}^{ - 1}} \otimes {{\bf{I}}_{\mid \Lambda \mid}}} \right)\left( {{\bf{D}}\Delta {{\bf{g}}_\Lambda } \otimes {{\bf{I}}_{\mid \Lambda \mid}}} \right){{\bf{x}}_\Lambda }  \\
     &=& {\bf{G}}\left( {{{\bf{D}}^{ - 1}} \otimes {{\bf{I}}_{\mid \Lambda \mid}}} \right)\left( {{{\bf{I}}_{\mid \Lambda \mid}} \otimes {{\bf{x}}_\Lambda }} \right){\bf{D}}\Delta {{\bf{g}}_\Lambda } \notag \\
     &=& {\bf{HD}}\Delta {{\bf{g}}_\Lambda }. \notag
\end{eqnarray}
When $\bf{W_x}$ is of full-row rank, the optimum problem (\ref{Equ:uCTLS}, \ref{Equ:Wx})
are equivalent to (\ref{Equ:xCTLS} - \ref{Equ:Wxinverse}), which
has been proved in \cite{ref:CTLS}.
\begin{equation}{\label{Equ:xCTLS}}
    {{\bf{x}}_{{\rm{CTLS}}}} = \mathop {\min }\limits_{{{\bf{x}}_\Lambda }} \left\| {{\bf{W}}_{\bf{x}}^\dag \left( {{\bf{y}} - {{\bf{\Phi }}_\Lambda }{{\bf{x}}_\Lambda }} \right)} \right\|_2^2
\end{equation}
\begin{equation}{\label{Equ:usolution}}
    \widehat {\bf{u}} = {\left. {{\bf{W}}_{\bf{x}}^\dag \left( {{\bf{y}} - {{\bf{\Phi }}_\Lambda }{{\bf{x}}_\Lambda }} \right)} \right|_{{{\bf{x}}_\Lambda } = {{\bf{x}}_{{\rm{CTLS}}}}}}
\end{equation}
\begin{equation}{\label{Equ:Wxinverse}}
    {\bf{W}}_{\bf{x}}^\dag  = {\bf{W}}_{\bf{x}}^{\rm{H}}{\left( {{{\bf{W}}_{\bf{x}}}{\bf{W}}_{\bf{x}}^{\rm{H}}} \right)^{ - 1}}
\end{equation}\par

It is quite difficult to obtain analytical solution to
(\ref{Equ:xCTLS}). A complex version of Newton method is developed
in \cite{ref:CTLS}, which is presented in Appendix A. Initial value of $\bf{x}_\Lambda$ required in Newton's method for (\ref{Equ:xCTLS}) can be given as:
\begin{equation}{\label{Equ:xini}}
    {{\bf{x}}_{{\rm{ini}}}} = {\bf{\Phi }}_\Lambda ^\dag {\bf{y}} = {\left( {{\bf{\Phi }}_\Lambda ^{\rm{H}}{{\bf{\Phi }}_\Lambda }} \right)^{ - 1}}{\bf{\Phi }}_\Lambda ^{\rm{H}}{\bf{y}}.
\end{equation}\par
${\widehat {\Delta {\bf{g}}}}_{\Lambda }$ is extracted from
$\hat{\bf{u}}$ via
$ {\widehat {\Delta {\bf{g}}}}_{\Lambda } = \left[ {{{\bf{D}}^{ - 1}}} \ \ {\bf{0_{\it{N}}}} \right]\widehat {\bf{u}}$, thus (\ref{Equ:cCTLS}) is solved.
The sketch of CTLS is given in Algorithm \ref{Alg:CTLS}.
As the authors' best knowledge, the convergence guarantees for this Newton method are still open question.

    \begin{table}[h]{\caption{The CTLS Technique}\label{Alg:CTLS}}
     \centering
     \begin{tabular}{l}\hline
         1) Input the dictionary $\bf{\Phi}_\Lambda$ and all the coefficient matrices ${\bf{R}}_i$. \\
         2) Compose $\bf{W}_{\bf{x}}$ and solve (\ref{Equ:xCTLS}) with the initial value given in (\ref{Equ:xini}). \\
         3) Calculate $\hat{\bf{u}}$ via (\ref{Equ:usolution}) and (\ref{Equ:Wxinverse}).   \\
         4) Extract ${\widehat {\Delta {\bf{g}}}}_{\Lambda }$  from $\hat{\bf{u}}$.     \\\hline
      \end{tabular}
    \end{table}

\subsection{Sketch of AMP-CTLS}{\label{subsec:sketch}}

Similarly to traditional MP methods \cite{ref:SP,ref:MP,ref:OMP1,ref:CoSaMP,ref:ROMP,ref:StOMP}, AMP-CTLS first greedily finds the support set. Then AMP-CTLS adaptively optimizes the grid points indexed in the support set. In this paper, we imitate the greedy approach of OMP, in which only one atom is added to the support set in each
iteration. If the number of atoms is known, terminate the iterations when the Cardinality of the support set reaches the pre-specified number. If it is not known, we can apply some other successfully used stopping criterions, e.g. norm of residual is below a threshold \cite{ref:residual}. A sketch of AMP-CTLS is presented in Algorithm \ref{Alg:AMP-CTLS}.
%    \begin{table}[!t]{\caption{the AMP-CTLS ALGORITHM}\label{Alg:AMP-CTLS}}
%         \begin{tabular*}{\columnwidth}{@{\extracolsep\fill}l}\hline
%              \tabincell{l}{1) divide the continuous parameter $f$ into grid point ${\hat{\bf{g}}}^{(0)}$; input the sparsity \\ level $M$. Set the support set ${\Lambda ^{\left( 0 \right)}} = \emptyset $, and the residual error ${{\bf{r}}^{\left( 0 \right)}} = {\bf{y}}$.}  \\
%              \tabincell{l}{2) calculate the correlations ${p_i} = \left\langle {{{\bf{r}}^{\left( k \right)}},{\bf{\Phi }}\left( {{{{\widehat g}_i}^{\left( k \right)}}} \right)} \right\rangle $.} \\
%              3) find the index $m = \mathop {\arg \max }\limits_i \left| {{p_i}} \right|$.\\
%              4) merge the support set ${\Lambda ^{(k+1)}} = {\Lambda^{(k)}} \cup \left\{ m \right\}$. \\
%              {\bf{5) solve (\ref{Equ:gridxlambda}) with the IJE algorithm.}}\\
%              6) update the residual error ${\bf{r}}^{(k+1)} = {\bf{y}} - {\bf{\Phi}}\left({\widehat{{\bf{g}_{\Lambda}}}}^{(k+1)} \right){\widehat{\bf{x}_{\Lambda}}}^{(k+1)}$.\\
%              7) increase $k$. Return to Step 2 if $k<M$.\\
%              \tabincell{l}{8) simultaneously output ${\widehat {{\bf{x}}}_{\Lambda }^{\left( {k} \right)}}$ and ${\widehat {{\bf{g}}}_{\Lambda }^{\left( {k} \right)}}$, and set the elements of \textbf{x} not \\ indexed in $\Lambda$ to 0.}\\ \hline
%          \end{tabular*}
%    \end{table}

    \begin{table}[h]{\caption{The AMP-CTLS Algorithm}\label{Alg:AMP-CTLS}}
        \centering
     \begin{tabular}{l}\hline
%         \begin{tabular*}{\columnwidth}{@{\extracolsep\fill}l}\hline
              \tabincell{l}{1) Divide the continuous parameter $f$ into grid point ${\hat{\bf{g}}}^{(0)}$; input the sparsity level $K$ or residual threshold $\delta$.\\ Set the support set ${\Lambda ^{\left( 0 \right)}} = \emptyset $, and the residual error ${{\bf{r}}^{\left( 0 \right)}} = {\bf{y}}$.}  \\
              \tabincell{l}{2) Calculate the correlations ${p_i} = \left\langle {{{\bf{r}}^{\left( k \right)}},{\bf{\Phi }}\left( {{{{\widehat g}_i}^{\left( k \right)}}} \right)} \right\rangle $.} \\
              3) Find the index $n = \mathop {\arg \max }\limits_i \left| {{p_i}} \right|$.\\
              4) Merge the support set ${\Lambda ^{(k+1)}} = {\Lambda^{(k)}} \cup \left\{ n \right\}$. \\
              {\bf{5) Solve (\ref{Equ:gridxlambda}) with the IJE algorithm.}} Then we get ${\widehat{{\bf{g}_{\Lambda}}}}^{(k+1)}$ and ${\widehat{\bf{x}_{\Lambda}}}^{(k+1)}$.\\
              6) Update the residual error ${\bf{r}}^{(k+1)} = {\bf{y}} - {\bf{\Phi}}\left({\widehat{{\bf{g}_{\Lambda}}}}^{(k+1)} \right){\widehat{\bf{x}_{\Lambda}}}^{(k+1)}$.\\
              7) Increase $k$. Return to Step 2 until stop criterion, e.g. $k<K$, $\|{\bf r}^{(k)}\|_2<\delta$ or $\|{\bf r}^{(k)}\|_2<\delta \|{\bf r}^{(k-1)}\|_2$, is satisfied.\\
              \tabincell{l}{8) Simultaneously output ${\widehat {{\bf{x}}}_{\Lambda }^{\left( {k} \right)}}$ and ${\widehat {{\bf{g}}}_{\Lambda }^{\left( {k} \right)}}$, and set the elements of \textbf{x} not indexed in $\Lambda$ to \textbf{0}.}\\ \hline
%          \end{tabular*}
          \end{tabular}
    \end{table}

\subsection{Convergence of the IJE Algorithm}{\label{subsec:convergence}}

Here, we analyze convergence of the IJE algorithm. Assume
that the mapping ${{\bf{g}}_\Lambda } \to {\bf{\Phi }}\left(
{{{\bf{g}}_\Lambda }} \right)$  is linear, which means
\begin{equation}{\label{Equ:linearmapping}}
    {\bf{\Phi }}\left( {{{\bf{g}}_\Lambda } + \Delta {{\bf{g}}_\Lambda }} \right) = {\bf{\Phi }}\left( {{{\bf{g}}_\Lambda }} \right) + {\bf{G}}  \left( {\Delta {{\bf{g}}_\Lambda } \otimes {{\bf{I}}_{\mid \Lambda \mid}}} \right),
\end{equation}
and {\bf{G}} should be a constant matrix.\\
{\textbf{Proposition}}. If the measurement \textbf{y} is perturbed by WGN and (\ref{Equ:linearmapping}) is obeyed, IJE monotonically reduces values of the penalty function in (\ref{Equ:gridxlambda}). The estimates of $\bf{x}_{\Lambda}$ and $\bf{g}_{\Lambda}$  satisfy:
\begin{equation}{\label{Equ:proposition}}
    \left\| {{\bf{y}} - {\bf{\Phi }}\big( {{{\widehat {\bf{g}}}_{\Lambda} {\left( {l} \right)}}} \big){{\widehat {\bf{x}}}_{\Lambda} {\left( {l} \right)}}} \right\|_2^2 \ge \left\| {{\bf{y}} - {\bf{\Phi }}\big( {{{\widehat {\bf{g}}}_{\Lambda} {\left( {l + 1} \right)}}} \big){{\widehat {\bf{x}}}_{\Lambda} {\left( {l + 1} \right)}}} \right\|_2^2.
\end{equation}
\textbf{Proof}.
Define a penalty function as follows:
\begin{eqnarray}{\label{Equ:newcctls}}
    f_{p}\left( {\Delta {{\bf{g}}_\Lambda },{{\bf{x}}_\Lambda }} \right) = \sigma _{\bf{w}}^2 \left\| {\bf{u}} \right\|_2^2 = \sigma _{\bf{w}}^2{\left( {\Delta {{\bf{g}}_\Lambda }} \right)^{\rm{H}}}{\bf{C}}_{{\bf{g}}}^{ - 1}\Delta {{\bf{g}}_\Lambda }
     +  \left\| {{\bf{y}} - {\bf{\Phi }}\big( {{{\widehat {\bf{g}}}_{\Lambda} {\left( {l} \right)}}} \big){{\bf{x}}_\Lambda } - {\bf{G}} \left( {\Delta {{\bf{g}}_\Lambda } \otimes {{\bf{I}}_{\mid \Lambda \mid}}} \right)}{{\bf{x}}_\Lambda } \right\|_2^2;
\end{eqnarray}
thus, ${{\widehat {\Delta {\bf{g}}}}_{\Lambda }}{\left( {l}
\right)}$ and ${\bf{x}}_{{\rm{CTLS}}}$ are obtained by solving
\begin{equation}{\label{Equ:cnewCTLS}}
    {{\widehat {\Delta {\bf{g}}}}_{\Lambda }{\left( {l}
    \right)}},{{\bf{x}}_{{\rm{CTLS}}}} = \mathop {\arg \min
    }\limits_{\Delta {{\bf{g}}_\Lambda },{{\bf{x}}_\Lambda }}
    {f_{p}}\left( {\Delta {{\bf{g}}_\Lambda },{{\bf{x}}_\Lambda }} \right),
\end{equation}
which is the same as (\ref{Equ:cCTLS}), for $\sigma_{\bf{w}}^2$ is a
constant. Thus, it is satisfied that
\begin{eqnarray}{\label{Equ:proof1}}
      f_{p}\left( {{{{\widehat {\Delta {\bf{g}}}}_{\Lambda }}{\left( {l} \right)}},{{\bf{x}}_{{\rm{CTLS}}}}}
     \right) \le f_{p}\big( {{\bf{0}},{{\widehat {\bf{x}}}_{\Lambda} {\left( {l} \right)}}}
     \big)
    = \left\| {{\bf{y}} - {\bf{\Phi }}\big( {{{\widehat {\bf{g}}}_{\Lambda} {\left( {l} \right)}}} \big){{\widehat {\bf{x}}}_{\Lambda} {\left( {l} \right)}}} \right\|_2^2.
\end{eqnarray}

Substitute ({\ref{Equ:updategrid}), (\ref{Equ:linearmapping}}) into $f_{p}\big( {{{{\widehat {\Delta {\bf{g}}}}_{\Lambda }}{\left( {l} \right)}},{{\bf{x}}_{{\rm{CTLS}}}}} \big)$, and note that ${\bf{C}}_{ \bf{g}}^{-1}$  is a positive
definite matrix; thus,
\begin{eqnarray}{\label{Equ:proof2}}
    f_{p}\left( {{{{\widehat {\Delta {\bf{g}}}}_{\Lambda }}{\left( {l} \right)}},{{\bf{x}}_{{\rm{CTLS}}}}} \right)
     &=& \left\| {{\bf{y}} - {\bf{\Phi }}\left( {{{\widehat {\bf{g}}}_{\Lambda} {\left( {l + 1} \right)}}} \right){{\bf{x}}_{{\rm{CTLS}}}}} \right\|_2^2
     + \sigma _{\bf{w}}^2{\left( {{{{\widehat {\Delta {\bf{g}}}}_{\Lambda }}{\left( {l} \right)}}} \right)^{\rm{H}}}{\bf{C}}_{ {\bf{g }}}^{ - 1}{{\widehat {\Delta {\bf{g}}}}_{\Lambda }{\left( {l} \right)}}\notag\\
     &\ge& \left\| {{\bf{y}} - {\bf{\Phi }}\left( {{{\widehat {\bf{g}}}_{\Lambda} {\left( {l + 1} \right)}}} \right){{\bf{x}}_{{\rm{CTLS}}}}} \right\|_2^2  \\
     &\ge& \left\| {{\bf{y}} - {\bf{\Phi }}\left( {{{\widehat {\bf{g}}}_{\Lambda} {\left( {l + 1} \right)}}} \right){{\widehat {\bf{x}}}_{\Lambda}{\left( {l + 1} \right)}}}
     \right\|_2^2, \notag
\end{eqnarray}
where the last inequality is taken from (\ref{Equ:LS}). The inequalities in (\ref{Equ:proof1}) and (\ref{Equ:proof2}) are transformed to equalities if and only if ${\widehat {\Delta {\bf{g}}}}_\Lambda {\left( l \right)} = {\bf{0}}$. \hfill $\square$ \par
For simplicity, we assume that the transform ${\bf \Phi} ({\bf g}_\Lambda)$ is linear. In some practical applications like harmonic retrieval, linearity is not strictly guaranteed. However, when atom mismatch $\Delta {\bf g}$ is significantly small, the higher order errors due to Taylor expansion (\ref{Equ:Taylor}) are ignorable, and (\ref{Equ:linearmapping}) is approximately satisfied. Numerical examples are performed in Section \ref{sec:sim} which demostrate the convergence of the proposed algorithm in the case of harmonic retrieval.
%    The IJE algorithm converges to a point ${{\bf{g}}}_{\Lambda} ^{\left(* \right)}$ if it exists, at which the solution of (\ref{Equ:cnewCTLS})
%    is ${\widehat {\Delta {\bf{g}}}}_\Lambda ^{\left( l \right)} = {\bf{0}}$.

\section{Application in the Harmonic Retrieval}{\label{sec:HarRet}}

In this section, we apply AMP-CTLS in harmonic retrieval. In Subsection \ref{subsec:HarmRet}, the signal model of harmonic retrieval is presented and adverse effects of MP approaches \cite{ref:SP,ref:MP,ref:OMP1,ref:CoSaMP,ref:ROMP,ref:StOMP} in harmonic retrieval is discussed. In Subsection \ref{subsec:HarmRetASR}, we detail the implementation of AMP-CTLS in harmonic retrieval.

\subsection{Signal Model of Harmonic Retrieval}{\label{subsec:HarmRet}}
Consider a complex sinusoidal signal
\begin{equation}{\label{Equ:HarmRtvSigModel}}
    {y_m} = \sum\limits_{k = 1}^K {{\alpha_k} \exp \left( {j2\pi {f_k}m} \right)}  + {w_m},
\end{equation}
where $y_m$ is the $m$th measurement, and $w_m$ is the $m$th noise, $m = 0,1,\dots,M-1$. There are $K$ sinusoids, and amplitude $\alpha_k$, frequency $f_k$ of the $k$th sinusoid are unknown parameters. When the sinusoids are sparse, i.e., $K \ll M$, harmonic retrieval problem can
be solved by grid-based CS approaches. Divide the digital
frequency $f \in \left[0 \ 1 \right)$
into $N$ grid points $\textbf{g}=[g_1,g_2,\dots,g_N]^{\rm T}$. When all frequencies are exactly at grid points, rewrite (\ref{Equ:HarmRtvSigModel}) as
\begin{equation}{\label{Equ:HMRMSum}}
    {y_m} = \sum\limits_{n = 1}^{N} {{x_n} \cdot \exp \left( {j2\pi {g_n}m} \right)}  + {w_m},
\end{equation}
where $g_n$ is the frequency of the $n$th grid point and
\begin{equation}{\label{Equ:HMRMxDef}}
    {x_n} = \left\{ \begin{array}{l}
    {\alpha _k}{\text{, the }}k{\text{th sinusoid is present at }}n{\text{th grid point,}}\\
    0{\text{, no sinusoid is present at }}n{\text{th grid point.}}
    \end{array} \right.
\end{equation}\par

Rewrite (\ref{Equ:HMRMSum}) in matrix form as
\begin{equation}{\label{Equ:HMRMMatrix}}
    {\bf{y}} = {\bf{\Phi x}} + {\bf{w}},
\end{equation}
where the $m$th-row, $n$th-column element of  $\bf \Phi$  is
of the form $\phi(m,n)=\exp(j2\pi g_{n}m)$. Apply CS methods to seek the sparsest solution of (\ref{Equ:HMRMMatrix}). Then, estimates of the frequencies and amplitudes are obtained with the indices and magnitudes of nonzero
coefficients in \textbf{x}, respectively. The sparsest solution can be obtained with computational MP methods, which greedily minimize the ${\ell _0}$ norm. It can also be obtained by minimizing the $\ell _1$ norm \cite{ref:harmonic1}, the quasi-norm \cite{ref:harmonic2,ref:harmonic3} or the $\ell _{p\le 1}$  p-norm-like diversity \cite{ref:harmonic4}.\par

We focus on MP methods in this paper for the high computation efficiency. However, conventional MP methods \cite{ref:SP,ref:MP,ref:OMP1,ref:CoSaMP,ref:ROMP,ref:StOMP}
suffer from performance degradation if the frequency space is not perfectly grided.
When the frequency is sparsely divided, sinusoids may lie off the grid points, and accuracy of frequency estimates is limited by the gap between neighboring grid points. MP methods iteratively search for the sinusoids. If an off-grid sinusoid emerges, the energy of this sinusoid can not be totally canceled and performs as an interference in the next iterations. The leakage of the energy may mask the weak sinusoids.
On the other hand, if the frequency space is densely divided, correlations between atoms are enhanced \cite{Tropp2004}, which also reduces the performance of MP methods. Especially in those MP methods that select multiple atoms into the support set in a single iteration, e.g. CoSaMP, SP, ROMP and StOMP, highly correlated atoms could be chosen in a same iteration, which impairs the numerical stability of projection onto the adopted atoms.\par

\subsection{Harmonic Retrieval with AMP-CTLS}{\label{subsec:HarmRetASR}}

The AMP-CTLS algorithm can be applied for harmonic retrieval. AMP-CTLS adaptively finds the atoms and recovers the sinusoids.
In those MP approaches with constant predefined atoms, frequency estimates are discrete values, depending on grid points. In AMP-CTLS frequency estimates are continuous, since estimates of the grid misalignments are continuous. In this subsection, we adjust two steps of AMP-CTLS presented in Section \ref{sec:SigMod} to better fit the harmonic retrieval problem.\par

Calculate the \textbf{R} matrix in (\ref{Equ:Rmatrix}). According to (\ref{Equ:Rmatrix}), the $m$th-row, $i$th-column element of the
${\bf{R}}_i$ is expressed as follows:
\begin{equation}
    {{\bf{R}}_i}\left( {m,i} \right){\bf{ = }}\exp \left( {j2\pi {g_i}m} \right) \cdot j2\pi m.
\end{equation}
Elements in other columns are all zeros.\par

Adjust the grid-updating formula in (\ref{Equ:updategrid}). In CTLS as presented in Subsection \ref{subsec:CTLS}, the grid misalignment $\Delta \bf{g}_{\Lambda}$ is assumed to be a complex vector; therefore, the estimate $\widehat{\Delta \bf{g}}_{\Lambda}$ is complex. However, frequency grid points are restrained to be real, so regularization $\Delta \bf{g}_{\Lambda} = (\Delta \bf{g}_{\Lambda})^*$  should be added to (\ref{Equ:uCTLS}) in the case of harmonic retrieval. Unfortunately, the solver becomes complex, which
is derived in Appendix B. %\ref{App:cvx}.
For simplicity, (\ref{Equ:updategrid}) is replaced with (\ref{Equ:updategridRe}) to approximatively update the grid points:
\begin{equation}{\label{Equ:updategridRe}}
   \widehat {\bf{g}}_\Lambda {\left( {l + 1} \right)} = {\widehat {\bf{g}}}_\Lambda {\left( {l} \right)} + {\mathop{\rm Re}\nolimits} \left( {{{{\widehat {\Delta {\bf{g}}}}_{\Lambda }}{\left( {l} \right)}}} \right).
\end{equation}

\section{Application in RSF Radar}{\label{sec:RSF}}

AMP-CTLS can also be applied in randomized step frequency (RSF)
radar. RSF radar can improve the range-velocity resolution and avoid range-velocity
coupling problems \cite{ref:analysisrsfr,ref:YiMin}.
However, RSF radar suffers from the sidelobe pedestal problem, which
results in small targets being masked by noise-like components due to dominant targets \cite{ref:YiMin}. Our problem of interest is to recover small targets. When the observed scene is sparse, i.e. only few targets exists, we can use sparse recovery to exploit the sparseness \cite{ref:SFR}. AMP-CTLS relieves the sidelobe pedestal problem in RSF radar and recovers small targets well.\par
Correlation-matrix-based spectral analysis methods, e.g. MUSIC, ESPRIT \cite{ref:stoica2005spectral}, are hard to directly utilized in range-Doppler estimation in RSF scheme. Since only one snapshot of radar data is available and radar echoes from different scatterers are coherent, smoothing technique is invoked to obtain a full rank correlation matrix \cite{Odendaal1994}. Smoothing method requires that the array is uniform and linear \cite{ref:stoica2005spectral}. However, in RSF radar, the echoes are determined by a random permutation of integers, see (\ref{Equ:RSFmodel2}); thus, the uniform and linear condition is not satisfied, which restricts application of correlation-matrix-based methods.\par
We discuss a specific example of RSF radar, in which the waveform is a monotone pulse signal and the frequency of the $m$th pulse is $f_{0}+C_{m}\delta f$, $m=0,1,\dots ,M-1$, where $f_0$ is
carrier frequency and $\delta f$ is frequency step size. $C_m$ is a random permutation of integers from 0 to $M-1$. The $m$th echo of radar can be expressed as (see \cite{ref:RSFR,ref:analysisrsfr,ref:YiMin}):
\begin{equation}{\label{Equ:RSFmodel}}
    {y_m} = \sum\limits_{k = 1}^K {{\alpha _k}{s_m}\left( {{p_k},{q_k}}
    \right)}  + {w_m},
\end{equation}
\begin{equation}{\label{Equ:RSFmodel2}}
   {s_m}\left( {p,q} \right) = \exp \Big( { - j2\pi {C_m}p - j2\pi m\left( {1 + {C_m}{{\delta f} \mathord{\left/
    {\vphantom {{\delta f} {{f_0}}}} \right.
    \kern-\nulldelimiterspace} {{f_0}}}} \right)q} \Big),
\end{equation}
where $w_m$ is noise in the $m$th echo. $K$ denotes the number of targets and $k$ denotes $k$th target. $\alpha _k$, $p_k$ and $q_k$ are to be learned. $\alpha _k$ presents the scattering intensity. ${p_k} \in \left[ 0 \ 1 \right)$ and ${q_k} \in \left[ 0 \ 1\right)$ are determined by range and radial velocity of the $k$th target, respectively. Note that in (\ref{Equ:RSFmodel2}) the echo is simultaneously related to the sequence $m$ and the random integer $C_m$.\par
Divide $p$ space into $C$ grid points ${p_c} = c/C$, $c = 0,1,\dots,C-1$. Divide $q$ space into $D$ grid points ${p_d} = d/D$, $d = 0,1,\dots,D-1$. Rewrite (\ref{Equ:RSFmodel}) as:
\begin{equation}{\label{Equ:RSFmatrix}}
    {\bf{y}} = {\bf{\Phi }}\left( {{\bf{p}},{\bf{q}}} \right){\bf{x}} + {\bf{w}},
\end{equation}
where the $m$th-row, $\left( c+dC \right)$th-column element of ${\bf{\Phi}}\left( {{\bf{p}},{\bf{q}}} \right) \in {\mathbb{C}}^{M \times CD}$ is ${s_m}\left( {{p_c},{q_d}} \right)$.
\par

AMP-CTLS is implemented to solve (\ref{Equ:RSFmatrix}).
First, we find the support set $\Lambda$ and then use IJE and CTLS to adjust the grid points, though CTLS described in Subsection \ref{subsec:CTLS} requires modification. The grid misalignment vector
consists of two parts: $p$ mismatch $\Delta {\bf{p}}
_{\Lambda} \in {\mathbb{R}}^{|\Lambda| \times 1}$ and $q$ mismatch $\Delta {\bf{q}}_{\Lambda} \in {\mathbb{R}}^{|\Lambda| \times 1}$,
$\Delta {\bf{g}}_{\Lambda}{\bf =}
\left[ \left(\Delta {\bf{p}}_{\Lambda} \right)^{\rm{T}} ,
\left(\Delta {\bf{q}}_{\Lambda} \right)^{\rm{T}}  \right]^{\rm{T}} \in {\mathbb{R}}^{2|\Lambda| \times 1}$.
The ${\bf{R}}\in {\mathbb{C}}^{M \times CD}$ matrix
\begin{equation}{\label{Equ:RSFRmatrix}}
{{\bf{R}}_{p_i}} = { {\frac{{\partial {\bf{\Phi }}\left(
{\bf{p}}_{\Lambda},{\bf{q}}_{\Lambda} \right)}}{{\partial {p_i}}}} },
{{\bf{R}}_{q_i}} = { {\frac{{\partial {\bf{\Phi }}\left(
{\bf{p}}_{\Lambda},{\bf{q}}_{\Lambda} \right)}}{{\partial {q_i}}}} }.
\end{equation}
Assume that $\Delta {\bf{p}} _{\Lambda}$ and $\Delta {\bf{q}}_{\Lambda}$ are independent of each other and of the noise.
The covariance matrix of $\Delta {\bf{g}} _{\Lambda}$ is
\begin{equation}{\label{Equ:RSFcov}}
{{\bf{C}}_{\bf{g}}} = \left[ {\begin{array}{*{20}{c}}
{{{\bf{C}}_{\bf{p}}}}&{\bf{0}}\\
{\bf{0}}&{{{\bf{C}}_{\bf{q}}}}
\end{array}} \right] \in {\mathbb{C}}^{2|\Lambda| \times 2|\Lambda|},
\end{equation}
where ${{\bf{C_p}}} = E\left[
{\Delta {{\bf{p}}_\Lambda }{{\left( {\Delta {{\bf{p}}_\Lambda }}
\right)}^{\rm{H}}}} \right] \in {\mathbb{C}}^{|\Lambda| \times |\Lambda|}$, ${{\bf{C_q}}} = E\left[
{\Delta {{\bf{q}}_\Lambda }{{\left( {\Delta {{\bf{q}}_\Lambda }}
\right)}^{\rm{H}}}} \right] \in {\mathbb{C}}^{|\Lambda| \times |\Lambda|}$,  ${\bf{C}}_{\bf{p}}^{ - 1} = {{\bf{D}}_{\bf p}^{\rm{H}}}{\bf{D_p}}$ and
${\bf{C}}_{\bf{q}}^{ - 1} = {{\bf{D}}_{\bf q}^{\rm{H}}}{\bf{D_q}}$.
In the case of RSF radar ${\bf u} = \left[ \left({\bf D_p}
\Delta {\bf{p}}_{\Lambda} \right)^{\rm{T}},
\left({\bf D_q} \Delta {\bf{q}}_{\Lambda} \right)^{\rm{T}},
\left( {\frac{1}{{{\sigma _{\bf{w}}}}}{\bf{w}}} \right)^{\rm{T}}
\right]^{\rm{T}}\in {\mathbb{C}}^{(2|\Lambda|+M) \times1}$, and ${{\bf{W}}_{\bf{x}}}=\left[
{\bf{H_p}},{\bf{H_q}},{{\sigma _{\bf{w}}}{{\bf{I}}_N}} \right] \in {\mathbb{C}}^{M \times (2|\Lambda|+M)}$,
where ${\bf{G}_p} = \left[
{{\bf{R}}_{p_1}},..,{{\bf{R}}_{p_{\mid \Lambda \mid}}} \right] \in {\mathbb{C}}^{M \times |\Lambda|^2}$,
${\bf{H_p}} = {\bf{G_p}}\left( {{{\bf{D}}_{\bf p}^{ - 1}} \otimes {{\bf{I}}_{\mid \Lambda \mid}}} \right)\left( {{{\bf{I}}_{\mid \Lambda \mid}} \otimes {{\bf{x}}_\Lambda }}
\right) \in {\mathbb{C}}^{M \times |\Lambda|}$,
${\bf{G}_q} = \left[
{{\bf{R}}_{q_1}},..,{{\bf{R}}_{q_{\mid \Lambda \mid}}} \right] \in {\mathbb{C}}^{M \times |\Lambda|^2}$
and
${\bf{H_q}} = {\bf{G_q}}\left( {{{\bf{D}}_{\bf q}^{ - 1}} \otimes {{\bf{I}}_{\mid \Lambda \mid}}} \right)\left( {{{\bf{I}}_{\mid \Lambda \mid}} \otimes {{\bf{x}}_\Lambda }}
\right) \in {\mathbb{C}}^{M \times |\Lambda|}$.
Since \textbf{p} and \textbf{q} are both real, formula (\ref{Equ:updategridRe}) is used to update grid points, in which the imaginary parts of $\Delta {\bf p}$ and $\Delta {\bf q}$ estimates are abandoned.

\section{Simulations}{\label{sec:sim}}

Numerical results are provided to illustrate the performance of the new algorithm. In all examples, the noise is additive Gaussian white noise.

\subsection{Accuracy of AMP-CTLS}{\label{subsec:SimAccuracy}}
We compare the accuracy of AMP-CTLS with standard OMP.
We assume that there is a single sinusoid in the measurements of form (\ref{Equ:HarmRtvSigModel}), where $\alpha=1$ and the signal to noise
ratio $\rm{SNR}={\alpha ^2/{\sigma ^2}} = 5$ dB, where $\sigma ^2$ is the variance of noise. The number of measurements $M$ is 32.
The frequency of sinusoid is varied between two adjoining frequency grid points.
The mean square errors (MSEs) of frequency estimates are calculated.
The MSEs are compared with the corresponding Cramer-Rao lower bound (CRB) \cite{Yau1992}.
The frequency is uniformly divided into $m$ grid points in xMP$_m$
(OMP$_M$, OMP$_{10M}$, CoSaMP$_M$, etc.) and into $M$ points in AMP-CTLS.
AMP-CTLS is configured as follows: IJE loops no more than 14
times; the normalization factors in (\ref{Equ:u}) are
${\bf{D}}={\bf{I}}/{\left(\sigma _{\Delta {\bf{f}}}\right)}$,
$\sigma _{\Delta {\bf{f}}}=0.005$ and $\sigma _{\bf{w}}=1$.
As shown in Fig. 1, MSEs of AMP-CTLS are
close to the CRB and lower than those of OMP, except
when the sinusoid is in the vicinity of the grid point.
\begin{figure}[h]
\centering
\includegraphics[width = 3in]{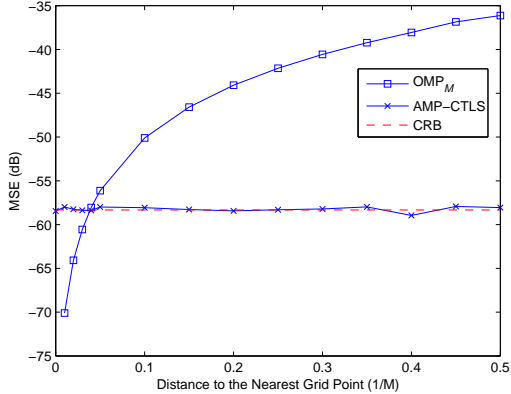}
\caption{MSEs of the frequency estimates obtained from 1000 independent Monte-Carlo trials via OMP and AMP-CTLS.
The distance to the nearest grid point is normalized by $1/M$.
CRB denotes Cramer-Rao bound.}
\label{Fig:acc}
\end{figure}

\subsection{Convergence of AMP-CTLS}{\label{subsec:SimIni}}
We first discuss the convergence speed of the proposed IJE algorithm in noise-free case. Suppose that the sinusoid is located at $f =9.5/M$, $M = 32$. Other conditions are the same as described in Subsection
\ref{subsec:SimAccuracy}. In the $l$-th iteration of IJE, we can obtain a grid point ${\hat g}(l)$ with (\ref{Equ:updategrid}) and residual error ${\bf r}(l)={ \bf y}-{\bf \Phi}\left({\hat g} \right){{\hat{x}(l)} }$ after (\ref{Equ:LS}). we calculate the norm of residual $\|{\bf r}(l)\|_2$ and the grid error $|{\hat g}(l)-f|$, and normalize the results with $\|{\bf r}(0)\|_2$ and $|{\hat g}(0)-f|$, respectively. As shown in Fig. 2, both the residual error and the grid error converge fast (about five steps) to 0 in noiseless case.\par
The purpose of what follows is to discuss the feasible zone of the initial grid points in noisy circumstance. In Section \ref{subsec:convergence}, the convergence analysis of IJE is based on the assumption that the transform $\Phi$ is linear. This is only approximately satisfied in harmonic retrieval when the higher order terms of Taylor expansion (\ref{Equ:Taylor}) is ignorable, which means that the grid points indexed in the support set are required to be close to the actual frequencies. We assign SNR $= 5$ dB and the initial frequency grid point as $g{(0)}=9/M$. The true frequency of the sinusoid varies from $9/M$ to $11/M$.
\begin{figure}
\centering
\includegraphics[width = 3in]{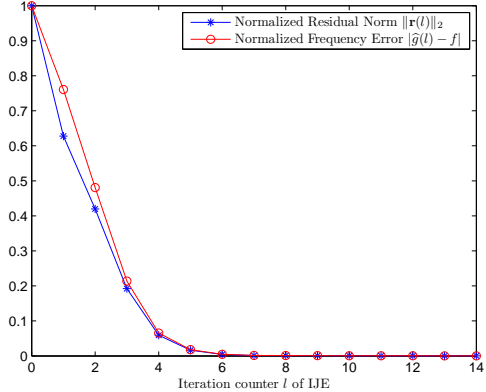}
\caption{The normalized errors at different IJE iterations in noiseless case. $\|{\bf r}(l)\|_2$ denotes the norm of residual error after $l$-th iteration. $f$ is actual frequency of the sinusoid and ${\hat g}(l)$ is the estimate of grid point. In the plots, $\|{\bf r}(l)\|_2$  and $|{\hat g}(l)-f|$ are normalize with $\|{\bf r}(0)\|_2$ and $|{\hat g}(0)-f|$, respectively.}
\label{Fig:convergence}
\end{figure}

As shown in Fig. 2, when the distance (normalized by $1/M$) between the true frequency and the initial grid point is less than 0.7, the initial grid is adjusted to be close to the actual value, and MSEs of the frequency estimates converge to CRB. When the distance is greater than 1, the AMP-CTLS curve is close to the initial distance, which means that AMP-CTLS fails to improve the initial grid, because errors of Taylor expansion cannot be ignored and affect convergence of the algorithm.

\begin{figure}[h]
\centering
\includegraphics[width = 3in]{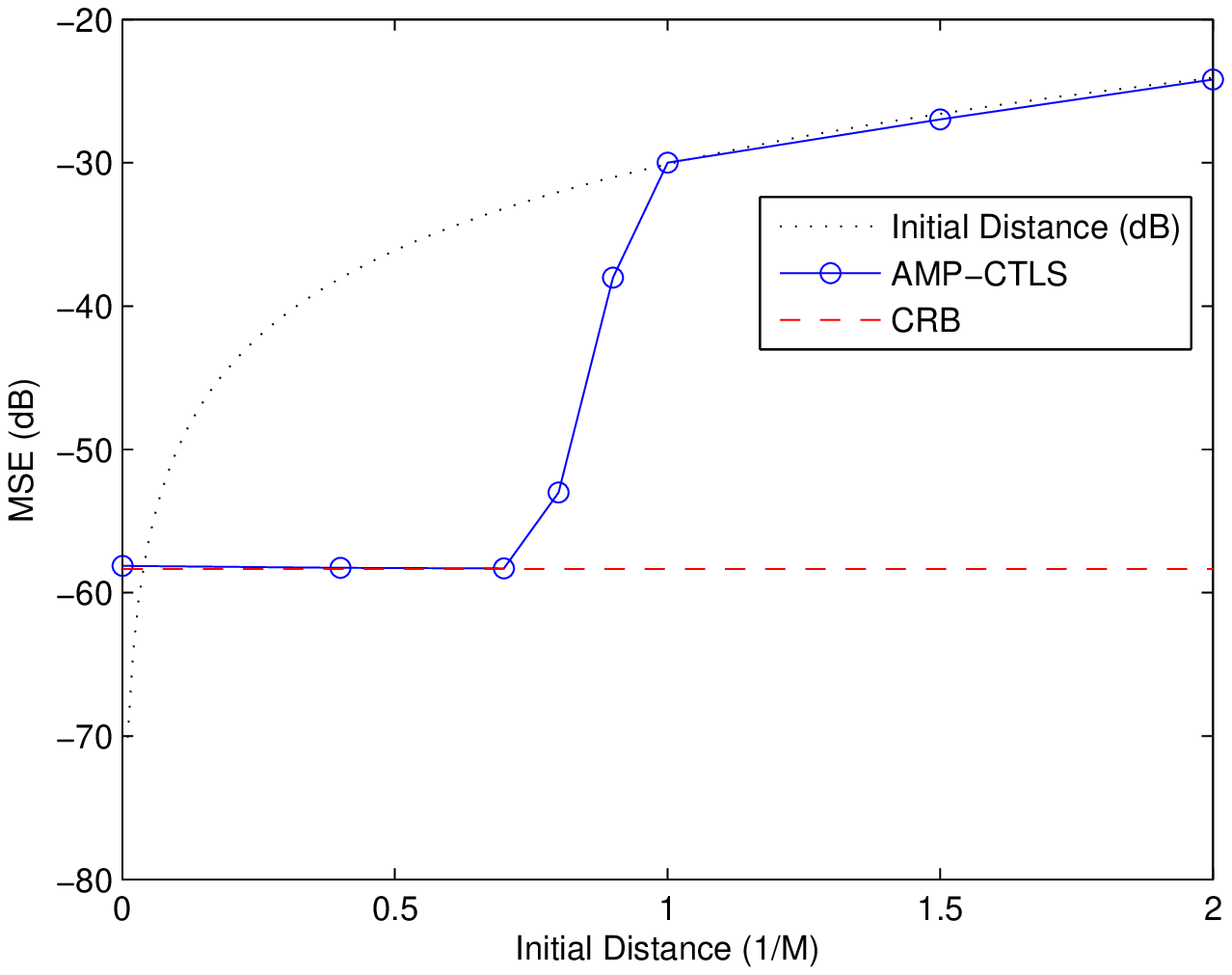}
\caption{MSEs of the frequency estimates versus distances
between the initial grid point and the actual values.
The distance is normalized by $1/M$.
The estimates are obtained from 1000 independent Monte-Carlo trials.
'Initial distance (dB)' in the legend denotes the square differences
between the frequency grid point and the true ones.}
\label{Fig:ini}
\end{figure}
\subsection{Input of Sparsity}{\label{subsec:number}}
In Subsection \ref{subsec:SimAccuracy} and \ref{subsec:SimIni}, we assume that the sparsity $K$, i. e. the number of modes, is known and we use $K$ to terminate AMP-CTLS, while a priori sparsity is not obligatory. When $K$ is unknown, we can use norm of residual error $ {\bf r}  =  \bf{y}-\bf{\Phi}\left( {\bf{g}_\Lambda} \right){{\bf{x}_\Lambda}}$ as termination criterion. Furthermore, AMP-CTLS does not seriously rely on the given sparsity $K'$, and the performance is slightly affected when $K'>K$. Suppose there are three sinusoids denoted as Si$_1$, Si$_2$, Si$_3$, where $\alpha _1=20$, $\alpha _2=15$, $\alpha _{3}=1$, $f_{1}=3.15/M$, $f_{2}=4.2/M$, $f_{3}=7.25/M$, $M = 32$. SNR$_{3} = {\alpha}_3^2/\sigma ^2 = 10$ dB. In AMP-CTLS, frequency is uniformly grided to $2M$ points, and other configurations are the same as described in Subsection
\ref{subsec:SimAccuracy}.\par
We calculate means of the final residual norm $\|{\bf r}^{(K')}\|_2$ versus $K'$ and present the results in Fig. 4. When all of the sinusoids have been chosen into the support set and $K' \ge K$, energy of the sinusoids are canceled thoroughly and only noise exist in the residual. The norm of residual error becomes small and is slowly reduced along with $K'$. The results illustrate that we can use threshold of values or decrease rate of the norm of residual to end AMP-CTLS loops.\par
\begin{figure}[h]
\centering
\includegraphics[width = 3in]{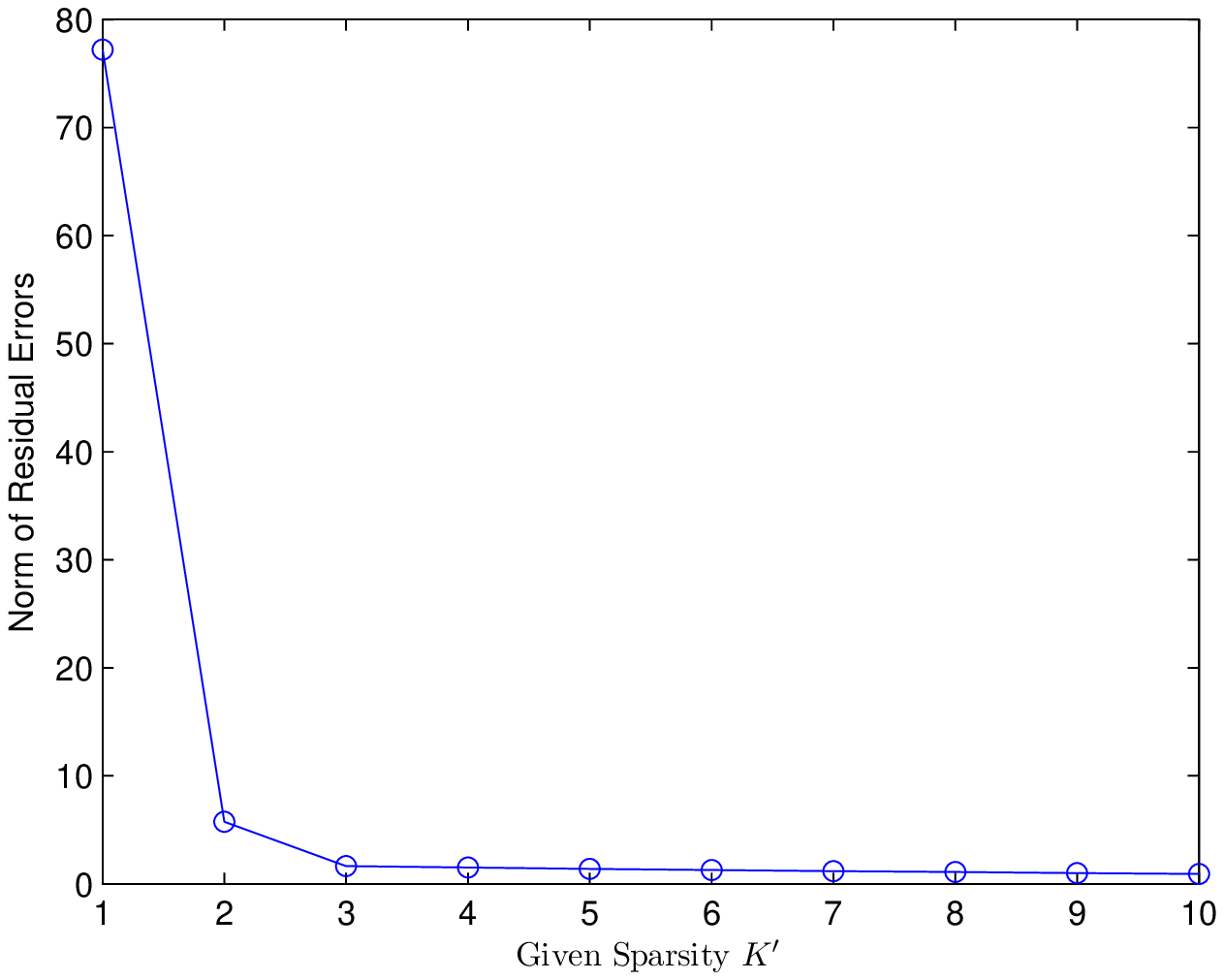}
\caption{Means of residual norm $\|{\bf r}^{(K')}\|_2$ versus given sparsity $K'$  obtained from 10000 independent Monte-Carlo trials.}
\label{Fig:normK}
\end{figure}
Spurious sinusoids emerge when $K'>K$. Denote the amplitude estimates by ${\hat{\alpha}_1},{\hat{\alpha}_2},\dots,{\hat{\alpha}_{K'}}$ in descend order of magnitudes and their counterparts of frequency estimates by ${\hat{f}_1},{\hat{f}_2},\dots,{\hat{f}_{K'}}$. MSEs of $f_3$ estimates versus $K'$ are presented in Fig. 5, which indicates that accuracy of frequency estimates of Si$_3$ is slightly affected ($<$ 2 dB) by $K'$.\par
\begin{figure}[!h]
\centering
\includegraphics[width = 3in]{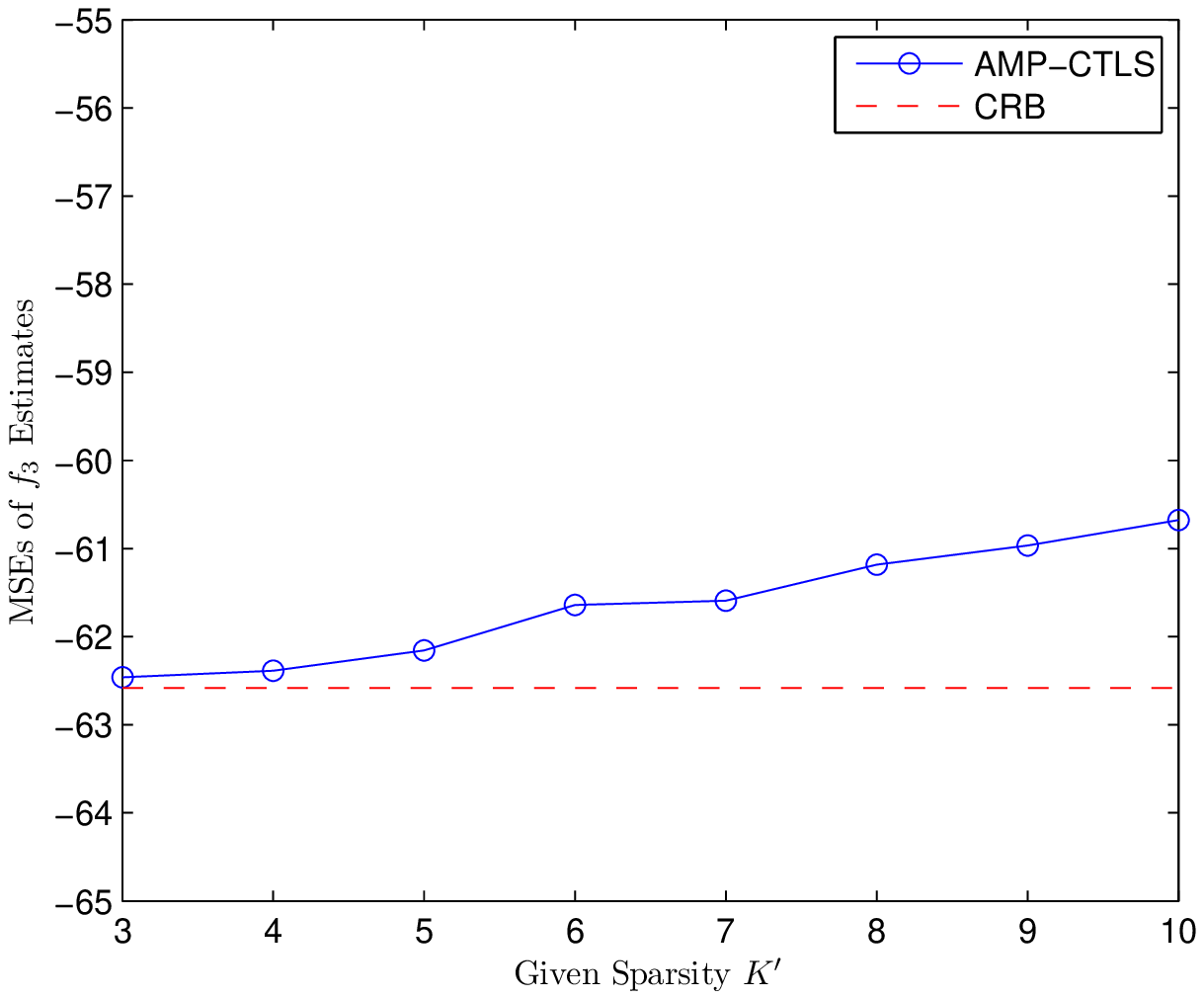}
\caption{MSEs of $f_3$ estimates versus given sparsity $K'$ obtained from 10000 independent Monte-Carlo trials. The results are compared with corresponding CRB.}
\label{Fig:MSEK}
\end{figure}
We also calculate $|{\hat{\alpha}_{K+1}}/{\alpha_K}|$ as measurement of the level of spurious sinusoids. Fig. 6 presents the results and shows that the ratios $|{\hat{\alpha}_{K+1}}/{\alpha_K}|$ stay at low level ($< 0.2$) and are not sensitive to $K'$.\par
\begin{figure}[h]
\centering
\includegraphics[width = 3in]{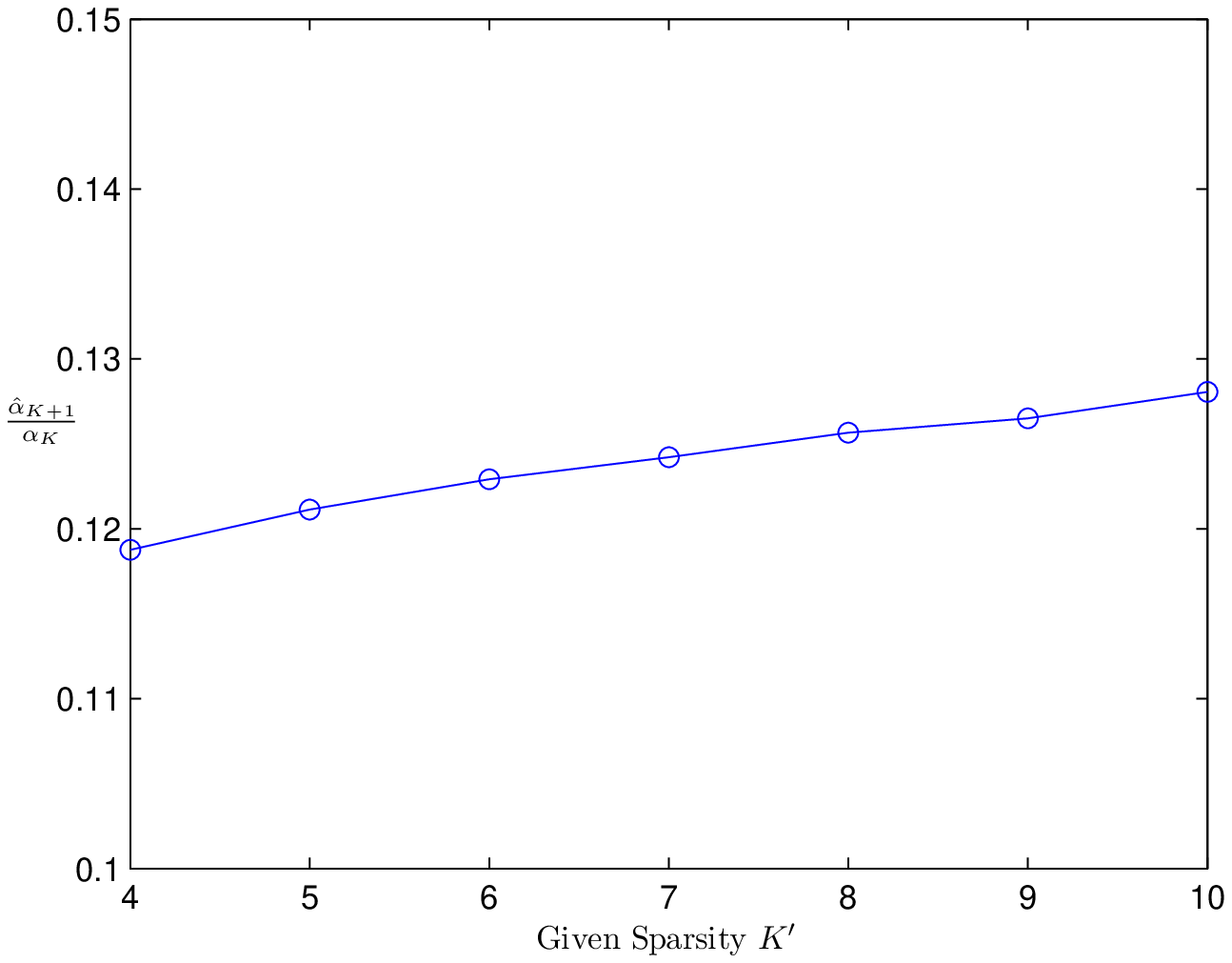}
\caption{Means of $|{\hat{\alpha}_{K+1}}/{\alpha_K}|$ versus given sparsity $K'$ obtained from 10000 independent Monte-Carlo trials, where ${\hat{\alpha}_{K+1}}$ denotes the biggest magnitude estimate of spurious sinusoid.}
\label{Fig:MSEK}
\end{figure}
Fig. 7 presents MSEs of $f_3$ estimates versus SNR at different $K'$. Noise variance $\sigma ^2$ is altered such that SNR$_{3}$ varies. The MSEs converges to CRB at high SNR (SNR$_{3}>2$ dB) when $K' = K = 3$. The results of $K' = 6$ are close to those of $K' = 3$.
\begin{figure}[h]
\centering
\includegraphics[width = 3in]{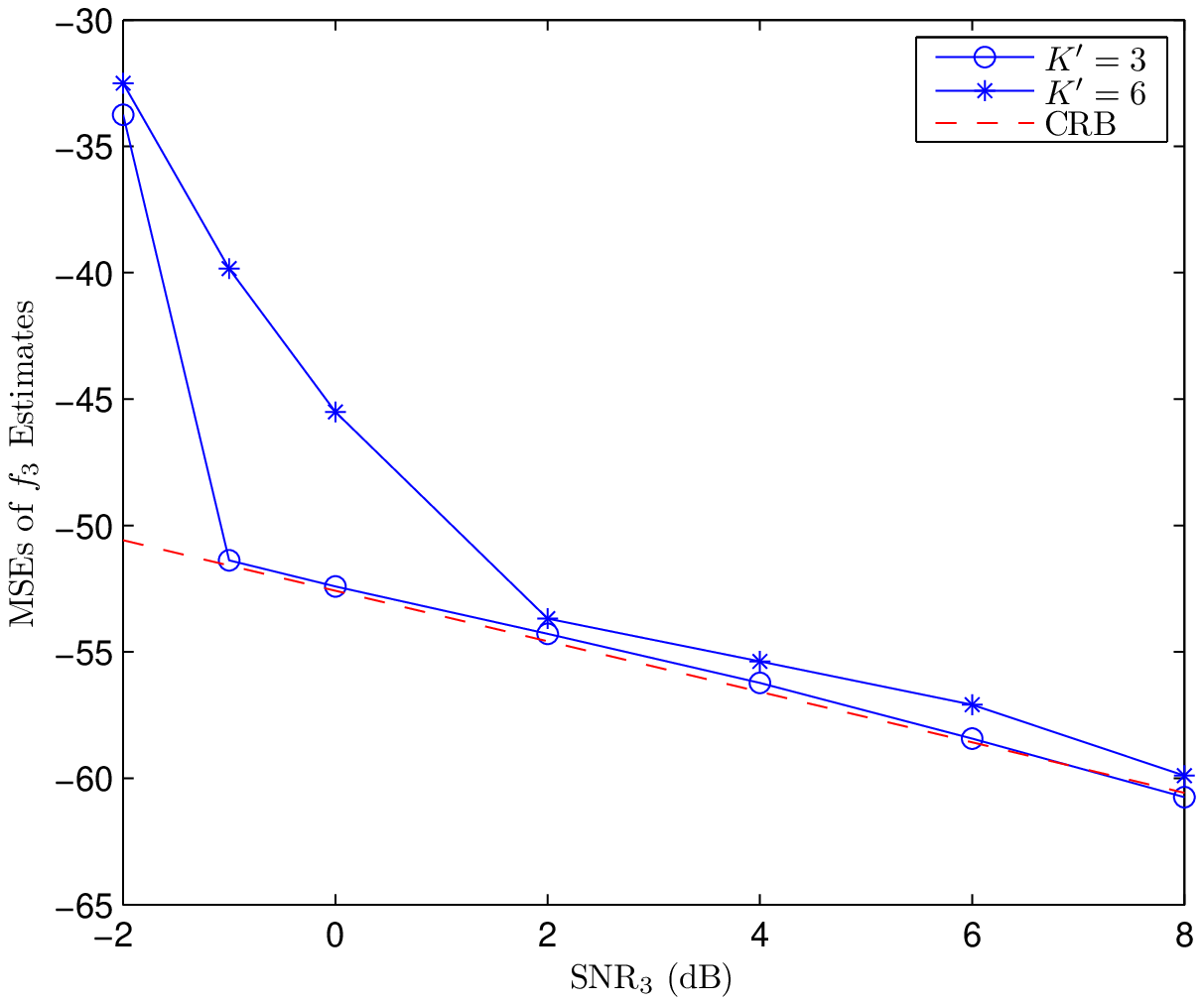}
\caption{MSEs of $f_3$ estimates versus SNR at different given sparsity $K'$ obtained from 1000 independent Monte-Carlo trials. The results are compared with CRB.}
\label{Fig:crbK}
\end{figure}

\subsection{Recovering Small Sinusoids}{\label{subsec:SimSmallSin}}
We compare the performance on recovering weak sinusoids of AMP-CTLS with CS methods, e.g. OMP and, CoSaMP, and conventional spectral analysis methods, e.g. ESPRIT and root MUSIC \cite{ref:stoica2005spectral}. Suppose there are three sinusoids denoted as Si$_1$, Si$_2$, Si$_3$, where $\alpha _1=20$, $\alpha _2=15$, $\alpha _{3}=1$, $f_{1}=3.15/M$, $f_{2}=5.2/M$, $f_{3}=3.95/M$, $M = 32$. SNR$_{3} = \alpha _{3}^2/\sigma ^2=5$ dB. The number of sinusoids $K = 3$ is assumed to be known. CoSaMP iterates 50 times. In both ESPRIT and root MUSIC, the model orders are set as $K$, and the covariance matrix orders are $M/2$ according to \cite{Mahata2004}. ESPRIT and root MUSIC output frequency estimates and the corresponding magnitudes are obtained by projection on these frequencies.  AMP-CTLS is configured the same as mentioned in Subsection \ref{subsec:number}.
%\begin{figure}[!h]
%\centering
%\subfloat[Comparison with OMP, CoSaMP] {\includegraphics[width=3in]{fig8a_3Sin}%
%\label{figa_3Sin}} \hfil
%\subfloat[Comparison with ESPRIT, root MUSIC] {\includegraphics[width=3in]{fig8b_3Sin}
%\label{figb_3Sin}}
%\caption{Amplitude and frequency estimates of the sinusoids obtained with OMP, CoSaMP, ESPRIT, root MUSIC and AMP-CTLS from 15 independent Monte-Carlo trials. The estimates are compared with the true parameters. Subscripts of OMP and CoSaMP denote numbers of grid points. Note that in OMP$_{100M}$, all of the sinusoids exactly lie on the grid points. The results of conventional MP methods and spectral analysis methods are respectively shown in (a) and (b) for readability.}
%\label{Fig:3Sin}
%\end{figure}
\begin{figure}[h]
\centering
\includegraphics[width = 5.7in]{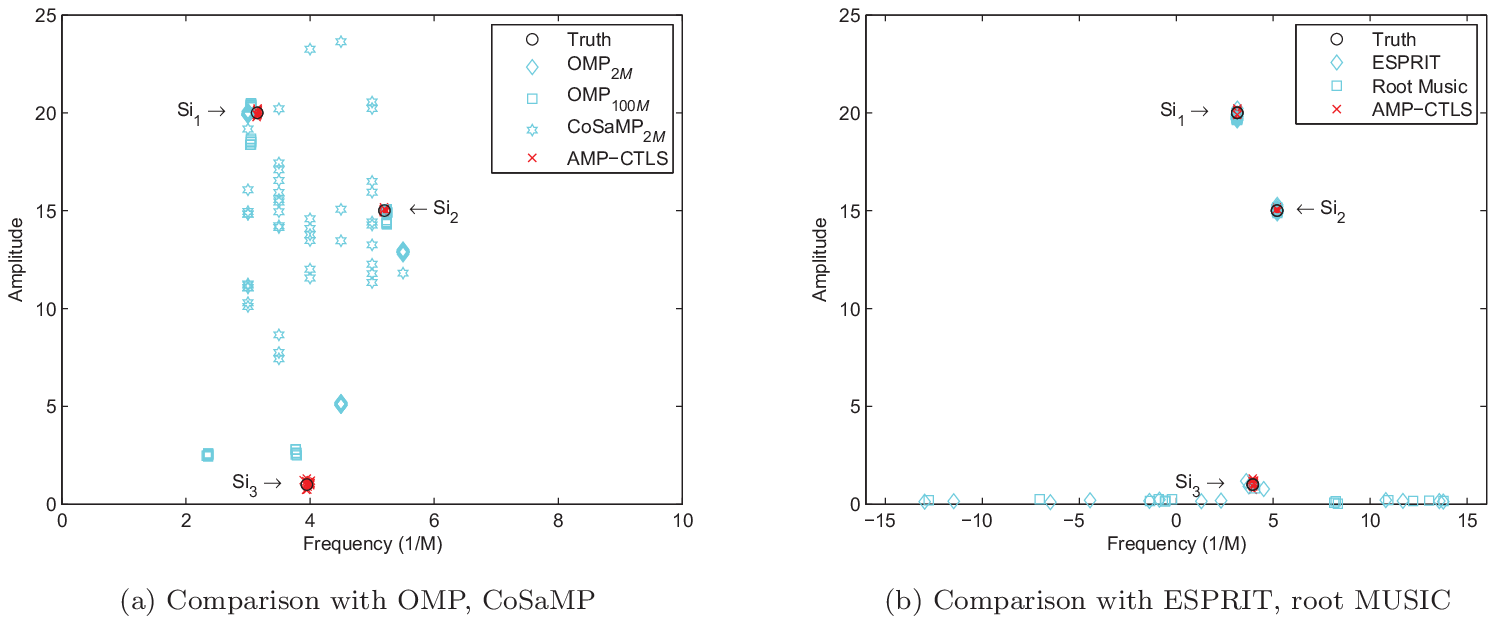}
\caption{Amplitude and frequency estimates of the sinusoids obtained with OMP, CoSaMP, ESPRIT, root MUSIC and AMP-CTLS from 15 independent Monte-Carlo trials. The estimates are compared with the true parameters. Subscripts of OMP and CoSaMP denote numbers of grid points. Note that in OMP$_{100M}$, all of the sinusoids exactly lie on the grid points. The results of conventional MP methods and spectral analysis methods are respectively shown in (a) and (b) for better readability.}
\label{Fig:3Sin}
\end{figure}

Some intuitive results are presented in Fig. 8. The Si$_3$ is recovered by the AMP-CTLS algorithm and is masked via other tested algorithms. CoSaMP$_{100M}$ is also tested, but the results are not displayed because the amplitude estimates are too large ($>1000$), which is caused by projection onto the ill-conditional matrix consisting of highly correlated atoms. In OMP$_{2M}$, the sinusoids are not exactly at the grid points, so the energies of Si$_1$ and Si$_2$ cannot be totally canceled in the beginning two iterations, and the leakage of the energies masks the smallest signal Si$_3$. In OMP$_{100M}$, all sinusoids are placed at the grid points, and Si$_1$ and Si$_2$ are better recovered than in OMP$_{2M}$, but energy leakage of dominant sinusoids still exists. In AMP-CTLS, the grid points are adaptively adjusted to match the sinusoids, so the algorithm is less sensitive to grid mismatch and can achieve better performance than OMP and CoSaMP even if the frequency space is sparsely divided. Since there is only one snapshot data, smoothing method \cite{Odendaal1994} is used in ESPRIT and root MUSIC to estimate the covariance matrix. This reduces the performance of frequency estimation.\par

%\subsection{Recovering Closely Spaced sinusoids} {\label{subsec:SimResolution}}
%
%We discuss the performance of AMP-CTLS and OMP$_{100N}$ in separating two close sinusoids. Suppose there are two sinusoids Si$_1$, Si$_2$. The parameters are given as $\alpha _{1}= \alpha _{2}=1$, $f_{1}=9/N$. $f_2$ is varied to change the separation between the sinusoids. The variance of the
%noise is -5 dB. AMP-CTLS is configured as discussed in
%the subsection \ref{subsec:SimAccuracy}. The MSEs of the frequency estimates versus the normalized separation are calculated.
%As shown in Fig. 5, AMP-CTLS has better harmonic super-resolution. The
%sinusoids interfere with each other when they
%are  located close together, and the correlations between the atoms
%and the measurement may introduce wrong indices into the support
%set. In OMP, an atom is never changed once it is chosen. In
%AMP-CTLS, the selected atoms are adaptively adjusted, and higher accuracy can be achieved.

\subsection{Range-Velocity Joint Estimate in RSF Radar}{\label{subsec:SimRSFR}}

In this subsection, we discuss merits of AMP-CTLS in recovering small targets with RSF radar. Suppose there are three targets: two large targets T$_1$ and T$_2$ and a small target T$_3$. The number of measurements $M$ is 32. The scattering intensities are $\alpha_{1}=\alpha_{2}=10$, $\alpha_{3}=1$, and the ranges and the velocities are set such that the $p$, $q$ parameters are
$p_{1}=10.1/M$, $p_{2}=10.7/M$, $p_{3}=20/M$, $q_{1}=19.4/M$, $q_{2}=10.2/M$ and $q_{3}=15.2/M$.
AMP-CTLS is configured as follows: the $p$, $q$ spaces are both
uniformly divided into $M$ grid points; the normalization
factors are ${\bf{D}_{\bf{p}}}={\bf{D}_{\bf{q}}}={{\bf{I}}} /{\sigma_{\Delta}}$,
$\sigma_{\Delta}=0.025$, $\sigma_{\bf{w}}=1$;
and the IJE algorithm iterates fewer than 14 times. In OMP$_m$,
both the $p$ and $q$ spaces are uniformly divided into $m$ grid points. Note that all of the targets lie on the
grid points in OMP$_{10M}$. We focus on the results of
recovering the weakest target T$_3$.
Change the noise covariance $\sigma ^2$; thus, the signal to noise ratio SNR$_{3}=\alpha _{3} ^2/{\sigma ^2}$ varies. Calculate
MSEs of $p_3$, $q_3$ parameters.
As shown in Fig. 9, the MSEs with AMP-CTLS are lower than those with OMP and converge to the CRB when the SNR$_3$ is no less than 2 dB. The difference between these MSEs of AMP-CTLS at high SNR and CRB is less than 0.5 dB.
%\begin{figure}[!h]
%\centering
%\subfloat[MSEs of $p_3$ estimates] {\includegraphics[width=3in]{fig9_RSFp}%
%\label{fig9_RSFp}} \hfil
%\subfloat[MSEs of $q_3$ estimates] {\includegraphics[width=3in]{fig9_RSFq}
%\label{fig9_RSFq}}
%\caption{MSEs of $p_3$ and $q_3$ estimates of OMP and AMP-CTLS versus SNR$_3$ obtained from 1000 independent Monto-Carlo trials. The results are compared with CRB.}
%\label{Fig:RSF}
%\end{figure}
\begin{figure}[h]
\centering
\includegraphics[width = 5.7in]{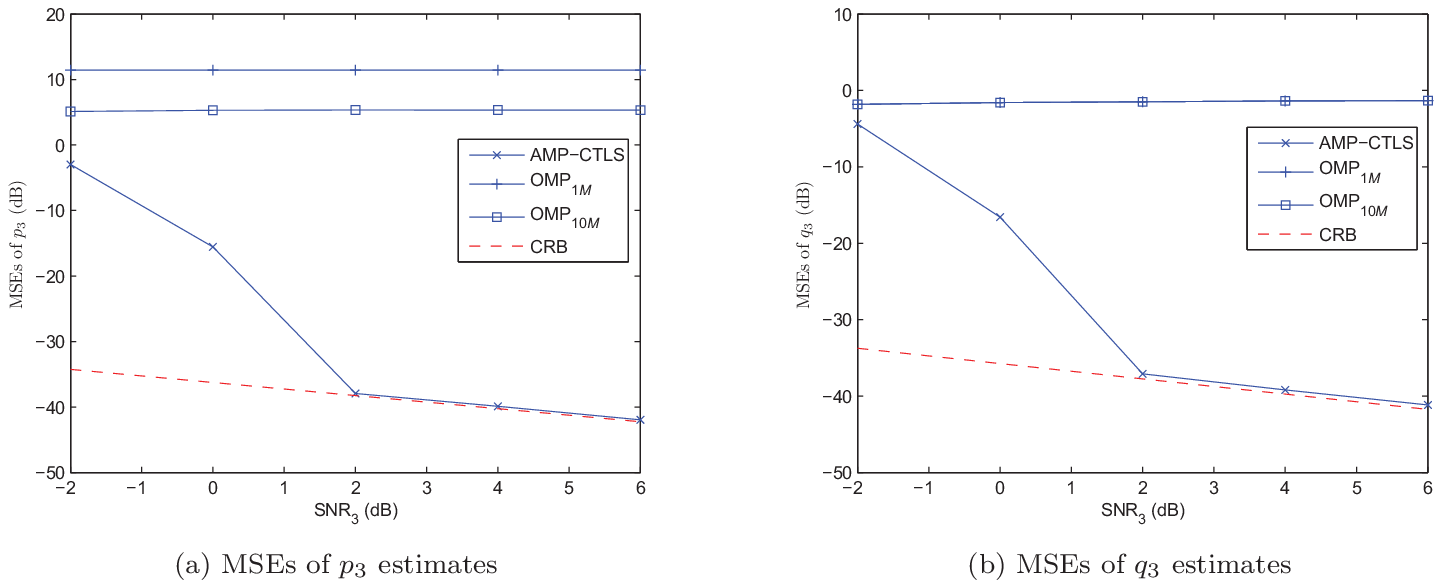}
\caption{MSEs of $p_3$ and $q_3$ estimates of OMP and AMP-CTLS versus SNR$_3$ obtained from 1000 independent Monto-Carlo trials. The results are compared with CRB.}
\label{Fig:RSF}
\end{figure}
\subsection{DoA Estimation}{\label{subsec:doa}}
In this subsection, AMP-CTLS is compared with the Lasso-based TLS method WSS-TLS \cite{ref:ZhuHao} on direction of arrival (DoA) estimation. The goal is estimating DoA of plane waves from far-field, narrowband sources with uniform linear array of antennas \cite{ref:ZhuHao}. We focus on the single-snapshot case. Suppose the antenna array contains $M = 8$ elements and the interval between neighboring elements $d = 1/2$ wavelength. There are two sources ($K = 2$) from angles $\theta_{1} = -29^{\circ}$ and $\theta_{2} = 13^{\circ}$. The amplitudes $\alpha_{1} = \alpha_{2} = 1$ and SNR$_{1}={\rm SNR}_{2}=\alpha_{1}^2/{\sigma^2}$, where $\sigma^2$ is the noise variance. The angle space from $-90^{\circ}$ to $90^{\circ}$ are uniformly divided to $N = 90$ grid points; thus both sources are $1^{\circ}$ off the nearest grid points. The WSS-TLS algorithm is set according to \cite{ref:ZhuHao}. Since WSS-TLS returns multiple nonzero DoA estimates, we choose two peaks with largest magnitudes as the estimates of $\theta_{1}$ and $\theta_{2}$. AMP-CTLS models DoA estimation as an harmonic retrieval problem and outputs frequency estimates ${\hat{f}}\in [0 \ 1)$. Denote ${\tilde{f}} = {\hat{f}}-0.5 \left({\rm{sgn}}({\hat{f}}-0.5)+1\right)$, where sgn$(\cdot)$ represents signum function; thus the DoA estimate with AMP-CTLS is obtained by ${\hat{\theta}} = \sin^{-1}({\tilde{f}}/d)$. In AMP-CTLS, IJE loops no more than 50 times; the normalization factors in (\ref{Equ:u}) are ${\bf{D}}={\bf{I}}/{\left(\sigma _{\Delta {\bf{f}}}\right)}$,
$\sigma _{\Delta {\bf{f}}}=0.005$, $\sigma _{\bf{w}}=1$. MSEs of $\theta_{1}$ and $\theta_{2}$ estimates versus SNR are shown in Fig. 10a and Fig. 10b respectively. The results indicate that MSEs of AMP-CTLS are closer to CRB than WSS-TLS.
%Besides, AMP-CTLS is based on greedy approach OMP. The advantage of OMP is that it admits simple, fast implementations \cite{Tropp2004}. We roughly test the computation load of AMP-CTLS and WSS-TLS. Set SNR$_{1}=$SNR$_{2}=15$ dB and run 100 independent Monto-Carlo trials. Matlab$^{\circledR}$ program for WSS-TLS takes 736 seconds on Intel$^{\circledR}$ Pentium$^{\circledR}$ processor T4400 (2.2GHz, 800MHz FSB), though Matlab$^{\circledR}$ implementation for AMP-CTLS takes 3 seconds on the same platform.
%\begin{figure}[!h]
%\centering
%\subfloat[MSEs of $\theta_{1}$ estimates] {\includegraphics[width=3in]{fig10a_DoA}%
%\label{fig10a_DoA}} \hfil
%\subfloat[MSEs of $\theta_{2}$ estimates] {\includegraphics[width=3in]{fig10b_DoA}
%\label{fig10b_DoA}}
%\caption{MSEs of $\theta_{1}$ and $\theta_{2}$ estimates of WSS-TLS and AMP-CTLS obtained from 1000 independent Monto-Carlo trials.}
%\label{Fig:DoA}
%\end{figure}
\begin{figure}[!h]
\centering
\includegraphics[width=5.5in]{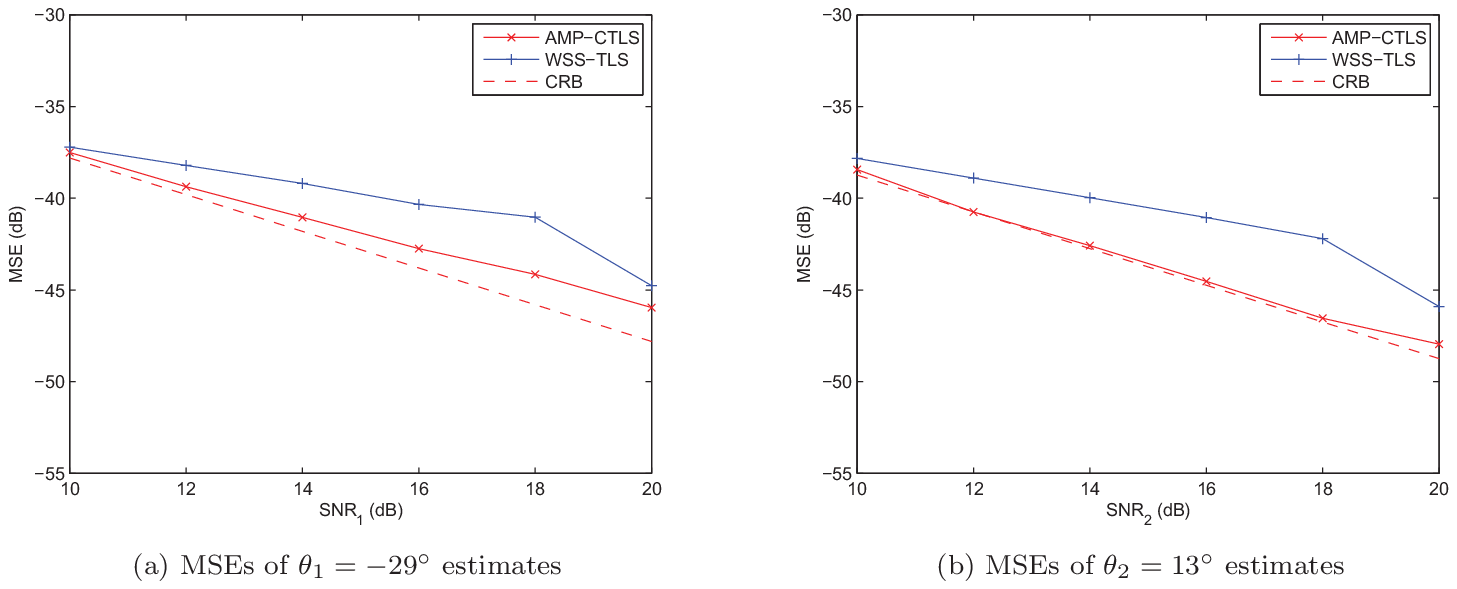}
\caption{MSEs of $\theta_{1}$ and $\theta_{2}$ estimates of WSS-TLS and AMP-CTLS obtained from 1000 independent Monto-Carlo trials.}
\label{Fig:DoA}
\end{figure}

\section{Conclusion}{\label{sec:conlusion}}
To alleviate the off-grid problem in grid-based MP methods,
we implement CTLS into the OMP framework and propose a new
algorithm, namely AMP-CTLS. Unlike traditional MP methods,
AMP-CTLS adaptively adjusts the grid and dictionary.
The convergence of the algorithm is analyzed.
Numerical examples indicate that the advantages of AMP-CTLS
over OMP and CoSaMP are twofold: 1) it is still efficient even when the continuous parameter space is sparsely divided, but OMP or CoSaMP suffers from performance degradation when the space is not divided
reasonably; 2) it can achieve higher accuracy and MSEs converge to CRB. In addition, the initial conditions of AMP-CTLS are discussed.

%\appendix
 \section*{Appendix A}{\label{App:CTLS}}
 We recall the complex version of Newton's method in \cite{ref:CTLS}.
 Minor changes are made because of differences in notation.
 The recursion formula of the complex Newton method for (\ref{Equ:xCTLS})
 is:
 \begin{equation}
 {\bf{x}}_{{\rm{CTLS}}}^{\left( {t + 1} \right)} = {\bf{x}}_{{\rm{CTLS}}}^{\left( t \right)} + {\left( {{{\bf{A}}^*}{{\bf{B}}^{ - 1}}{\bf{A}} - {{\bf{B}}^*}} \right)^{ - 1}}\left( {{{\bf{a}}^*} - {{\bf{A}}^*}{{\bf{B}}^{ - 1}}{\bf{a}}} \right),
 \end{equation}
 where
 \begin{equation}
{\bf{h}} = {\left( {{{\bf{W}}_{\bf{x}}}{\bf{W}}_{\bf{x}}^{\rm{H}}} \right)^{ - 1}}\left( {{{\bf{\Phi }}_\Lambda }{\bf{x}}_{{\rm{CTLS}}}^{\left( t \right)} - {\bf{y}}} \right),
 \end{equation}
  \begin{equation}
 \widehat {\bf{u}} =  - {\bf{W}}_{\bf{x}}^{\rm{H}}{\bf{h}},
 \end{equation}
  \begin{equation}
\widetilde {\bf{B}} = {{\bf{\Phi }}_\Lambda } + {\bf{G}}\left(
{{\bf{D}}^{{\rm{ - 1}}} \otimes {{\bf{I}}_{\mid \Lambda \mid }}}
\right)\left( {{\bf I}_{{|\Lambda|} \times
(M+|\Lambda|)} \hat{\bf u} \otimes {{\bf{I}}_{\mid \Lambda
\mid }}} \right),
\end{equation}
  \begin{equation}
{{\bf{Q}}_j} = \left[ {\left[ {{{\left( {{{\bf{R}}_1}}
\right)}_{\left\{ j \right\}}},\dots,{{\left( {{{\bf{R}}_{\mid
\Lambda \mid }}} \right)}_{\left\{ j \right\}}}}
\right]{\bf{D}}^{{\rm{ - 1}}},{{\bf{0}}_M}} \right],j \le {\mid
\Lambda \mid },
\end{equation}
  \begin{equation}
\widetilde {\bf{Q}} = \left[
{{\bf{Q}}_1^{\rm{H}}{\bf{h}},{\bf{Q}}_2^{\rm{H}}{\bf{h}},\dots,{\bf{Q}}_{\mid
\Lambda \mid }^{\rm{H}}{\bf{h}}} \right], \end{equation}
  \begin{equation}
{\bf{a}} = {\left( {{{\bf{h}}^{\rm{H}}}\widetilde {\bf{B}}} \right)^{\rm{T}}}, \end{equation}
  \begin{equation}
{\bf{A}} =  - {\widetilde {\bf{Q}}^{\rm{H}}}{\bf{W}}_{\bf{x}}^\dag
\widetilde {\bf{B}} - {\left( {{{\widetilde
{\bf{Q}}}^{\rm{H}}}{\bf{W}}_{\bf{x}}^\dag \widetilde {\bf{B}}}
\right)^{\rm{T}}} ,\end{equation}
  \begin{equation}
{\bf{B}} = {\widetilde {\bf{Q}}^{\rm{H}}}\left(
{{\bf{W}}_{\bf{x}}^\dag {{\bf{W}}_{\bf{x}}} - {\bf{I}}}
\right)\widetilde {\bf{Q}} + {\left( {{{\widetilde
{\bf{B}}}^{\rm{H}}}{{\left(
{{{\bf{W}}_{\bf{x}}}{\bf{W}}_{\bf{x}}^{\rm{H}}} \right)}^{ -
1}}\widetilde {\bf{B}}} \right)^{\rm{T}}}. \end{equation}

 \section*{Appendix B}
 When the frequencies are constrained to be real,
 the CTLS solver becomes more complex.
 Some notations are introduced for simplicity as follows: ${\bf W}_{1}={\bf H}$,
 ${{\bf{W}}_2} = {\sigma _{\bf{w}}}{{\bf{I}}_N}$,
 ${\bf{u}} = {\left[ {{\bf{u}}_1^{\rm{T}},{\bf{u}}_2^{\rm{T}}} \right]^{\rm{T}}}$,
 ${\bf{z}} = {\bf{y}} - {{\bf{\Phi }}_\Lambda }{{\bf{x}}_\Lambda
 }$. Notice that the matrix $\bf H$ is relative to ${\bf
 x}_{\Lambda}$.
 Replace the optimum problem in (\ref{Equ:uCTLS}), (\ref{Equ:Wx}) with:
  \begin{equation}{\label{Equ:}}
\widehat {\bf{u}},{{\bf{x}}_{{\rm{CTLS}}}} =\mathop {\arg \min
}\limits_{{\bf{u}}_1,{\bf{u}}_2,{{\bf{x}}_\Lambda }}
{\bf{u}}_1^{\rm{T}}{{\bf{u}}_1} + {\bf{u}}_2^{\rm{H}}{{\bf{u}}_2},
 \end{equation}
  \begin{equation}
  s.t. \left\{ \begin{array}{l}
{\bf{z}} - \left[ {{{\bf{W}}_1},{{\bf{W}}_2}} \right]\left[ \begin{array}{l}
{{\bf{u}}_1}\\
{{\bf{u}}_2}
\end{array} \right] = {\bf{0}}\\
{\bf{u}}_1^* = {{\bf{u}}_1}
\end{array} \right.
 .\end{equation}
First, suppose ${\bf x}_{\Lambda}$ is known, and seek the solution
of \textbf{u}. If both ${\bf W}_1$ and
 ${\bf W}_2$ are of full-row rank, we have
  \begin{equation}{\label{Equ:AppBsolutionstart}}
  {{\bf{u}}_1} = 2{\mathop{\rm Re}\nolimits} \left( {{\bf{W}}_1^{\rm{H}}{\bf{v}}} \right),
 \end{equation}
  \begin{equation}
  {{\bf{u}}_2} = 2{\bf{W}}_2^{\rm{H}}{\bf{v}},
 \end{equation}
 where
  \begin{equation}{\label{Equ:AppBsolutionmid}}
  {\bf{v}} =  - {\left( {{\bf{C}}_2^{ - 1}{{\bf{C}}_1} - {{\left( {{\bf{C}}_1^{ - 1}{{\bf{C}}_2}} \right)}^*}} \right)^{ - 1}}\left( {{\bf{C}}_2^{ - 1}{\bf{z}} - {{\left( {{\bf{C}}_1^{ - 1}{\bf{z}}} \right)}^*}} \right),
 \end{equation}
  \begin{equation}
  {{\bf{C}}_1} = {{\bf{W}}_1}{\bf{W}}_1^{\rm{H}} + 2{{\bf{W}}_2}{\bf{W}}_2^{\rm{H}},
 \end{equation}
  \begin{equation}{\label{Equ:AppBsolutionend}}
  {{\bf{C}}_2} = {{\bf{W}}_1}{\bf{W}}_1^{\rm{T}}.
 \end{equation}
 The solution ${\bf{u}}_1$, ${\bf{u}}_2$ depends on ${\bf
 x}_{\Lambda}$. Then, calculate ${\bf x}_{\Lambda}$ as
 \begin{equation}{\label{Equ:xrealu}}
{{\bf{x}}_{{\rm{CTLS}}}} =\mathop {\arg \min }\limits_{{\bf
 x}_{\Lambda}}{\bf{u}}_1^{\rm{T}}{{\bf{u}}_1} + {\bf{u}}_2^{\rm{H}}{{\bf{u}}_2}.
 \end{equation}
 \par
However, it is difficult to solve ({\ref{Equ:xrealu}}), because the
Jacobi matrix and Hessian matrix of ${\bf{u}}_1^{\rm{T}}{{\bf{u}}_1}
+ {\bf{u}}_2^{\rm{H}}{{\bf{u}}_2}$
 versus ${\bf x}_{\Lambda}$ are complex.
 In this paper, we simply consider the frequencies to be complex
 and ignore the imaginary parts.

 \textbf{Proof}:
We prove that when ${\bf x}_{ \Lambda}$ is known, the solution of
 \textbf{u} is given as ({\ref{Equ:AppBsolutionstart}}) to ({\ref{Equ:AppBsolutionend}}).
  The Lagrangian
  \begin{equation}
L\left( {{\bf{u}},{\bf{v}}} \right) = \frac{1}{2}{\bf{u}}_1^{\rm{T}}{{\bf{u}}_1}  - {\bf{v}}_{}^{\rm{H}}\left( {{\bf{z}} + {{\bf{W}}_1}{{\bf{u}}_1} + {{\bf{W}}_2}{{\bf{u}}_2}} \right)
 + \frac{1}{2}{\bf{u}}_2^{\rm{H}}{{\bf{u}}_2} - {\bf{v}}_{}^{\rm{T}}\left( {{{\bf{z}}^*} + {\bf{W}}_1^*{{\bf{u}}_1} + {\bf{W}}_2^*{\bf{u}}_2^*} \right)
 \end{equation}
 can be expressed as follows:
  \begin{eqnarray}
L\left( {{\bf{u}},{\bf{v}}} \right) &=& \frac{1}{2}{\left( {{{\bf{u}}_1} - 2{\mathop{\rm Re}\nolimits} \left( {{\bf{W}}_1^{\rm{H}}{\bf{v}}} \right)} \right)^{\rm{T}}}\left( {{{\bf{u}}_1} - 2{\mathop{\rm Re}\nolimits} \left( {{\bf{W}}_1^{\rm{H}}{\bf{v}}} \right)} \right)
 - 2{\left( {{\mathop{\rm Re}\nolimits} \left( {{\bf{W}}_1^{\rm{H}}{\bf{v}}} \right)} \right)^{\rm{T}}}{\mathop{\rm Re}\nolimits} \left( {{\bf{W}}_1^{\rm{H}}{\bf{v}}} \right) \notag \\
 &-& 2{{\bf{v}}^{\rm{H}}}{{\bf{W}}_2}{\bf{W}}_2^{\rm{H}}{\bf{v}} + \frac{1}{2}{\left( {{{\bf{u}}_2} - 2{\bf{W}}_2^{\rm{H}}{\bf{v}}} \right)^{\rm{H}}}\left( {{{\bf{u}}_2} - 2{\bf{W}}_2^{\rm{H}}{\bf{v}}} \right) - 2{\mathop{\rm Re}\nolimits} \left[ {{{\bf{v}}^{\rm{H}}}{\bf{z}}} \right].
 \end{eqnarray}
 When ${{\bf{u}}_1} = 2{\mathop{\rm Re}\nolimits} \left( {{\bf{W}}_1^{\rm{H}}{\bf{v}}} \right)$
 and  ${{\bf{u}}_2} = 2{\bf{W}}_2^{\rm{H}}{\bf{v}}$,
 the Lagrangian reaches the infimum; thus, the Lagrange dual function is obtained as
  \begin{equation}
\gamma \left( {\bf{v}} \right) = \mathop {\inf }\limits_{\bf{u}}
L\left( {{\bf{u}},{\bf{v}}} \right)
 =   - 4{{\bf{v}}^{\rm{T}}}{\bf{W}}_2^*{\bf{W}}_2^{\rm{T}}{{\bf{v}}^*} - 4{\mathop{\rm Re}\nolimits} \left( {{{\bf{v}}^{\rm{H}}}{\bf{z}}} \right)
 - \left( {{{\bf{v}}^{\rm{H}}}{{\bf{W}}_1} + {{\bf{v}}^{\rm{T}}}{\bf{W}}_1^*} \right)\left( {{\bf{W}}_1^{\rm{H}}{\bf{v}} + {\bf{W}}_1^{\rm{T}}{{\bf{v}}^*}} \right).
 \end{equation}
 Calculate the Jacobi and Hessian matrix of $\gamma({\bf v})$:
  \begin{equation}
  \frac{{\partial \gamma }}{{\partial {{\bf{v}}^*}}} =  - \frac{1}{2}{{\bf{W}}_1}{\bf{W}}_1^{\rm{H}}{\bf{v}} - \frac{1}{2}{{\bf{W}}_1}{\bf{W}}_1^{\rm{T}}{{\bf{v}}^*} - {{\bf{W}}_2}{\bf{W}}_2^{\rm{H}}{\bf{v}} - \frac{1}{2}{\bf{z}},
 \end{equation}
  \begin{equation}
  \frac{{{\partial ^2}\gamma }}{{\partial {{\bf{v}}^{\rm{T}}}\partial {{\bf{v}}^*}}} =  - \frac{1}{2}{{\bf{W}}_1}{\bf{W}}_1^{\rm{H}} - {{\bf{W}}_2}{\bf{W}}_2^{\rm{H}}.
 \end{equation}
 Because ${\bf W}_1$, ${\bf W}_2$  are of full-row rank,
 $\frac{{{\partial ^2}\gamma }}{{\partial {{\bf{v}}^{\rm{T}}}\partial {{\bf{v}}^*}}}$
 is a negative definite matrix.
 We solve \textbf{v} with the optimum condition
 $\frac{{\partial \theta }}{{\partial {{\bf{v}}^*}}} = {\bf{0}}$,
 and obtain (\ref{Equ:AppBsolutionmid}) to (\ref{Equ:AppBsolutionend}).
 The proof is complete.

%%%%%%%%%%%%%%%%%%%%%%%%%%%%%%%%%
%\section*{Author's contributions}
%    Text for this section \ldots

%%%%%%%%%%%%%%%%%%%%%%%%%%%
\section*{Acknowledgements}
  \ifthenelse{\boolean{publ}}{\small}{}
  This work was supported in part by the National Natural Science
Foundation of China (No. 40901157) and in part by the National Basic
Research Program of China (973 Program, No. 2010CB731901).\par
Thanks to the anonymous reviewer for many valuable comments and to Hao Zhu for helpful discussions and her Matlab$^{\circledR}$ programs of WSS-TLS.

%%  Bmc_article.bst  will be used to                       %%
%%  create a .BBL file for submission, which includes      %%
%%  XML structured for BMC.                                %%
%%  After submission of the .TEX file,                     %%
%%  you will be prompted to submit your .BBL file.         %%

%\newpage
%{\ifthenelse{\boolean{publ}}{\footnotesize}{\small}
% \bibliographystyle{bmc_article}  % Style BST file
%  \bibliography{bmc_article} }     % Bibliography file (usually '*.bib' )
%\bibliographystyle{IEEEtran}
\bibliography{ompctls}

\end{bmcformat}
\end{document}